\documentclass[letterpaper,twocolumn,10pt]{article}
\usepackage{usenix}

\usepackage{tikz}
\usepackage{amsmath}
\usepackage{amssymb}
\usepackage{filecontents}
\usepackage{booktabs} 
\usepackage{multirow}
\usepackage{algorithm}
\usepackage{algorithmic}
\usepackage{newfloat}
\usepackage{listings}
\usepackage{colortbl}
\usepackage{enumitem}
\usepackage{amsmath, amssymb}
\usepackage[dvipsnames]{xcolor} 
\usepackage[most]{tcolorbox}
\usepackage{xcolor}
\usepackage{graphicx}
\usepackage{subcaption} 
\usepackage{diagbox}
\usepackage[available]{usenixbadges}

\definecolor{MyBlueGreen}{rgb}{0.0, 0.87, 0.87}

\newcommand{\eg}{\textit{e.g.}}

\begin{document}

\date{}

\title{\Large \bf Sirens' Whisper: Inaudible Near-Ultrasonic Jailbreaks of Speech-Driven LLMs}


\author{
Zijian Ling$^{12*\ddagger}$,
Pingyi Hu$^{1*\ddagger}$,
Xiuyong Gao$^{1*\ddagger}$,
Xiaojing Ma$^{1\dagger\ddagger}$,
Man Zhou$^{12\dagger\ddagger}$, 
Jun Feng$^{1\ddagger}$,\\ Songfeng Lu$^{1\ddagger}$,
Dongmei Zhang$^{3}$, Bin Benjamin Zhu$^{3}$\\
$^{1}$School of Cyber Science and Engineering, Huazhong University of Science and Technology \\
$^{2}$State Key Laboratory of Internet Architecture, Tsinghua University\\ 
$^{3}$Microsoft Corporation\\
Email: \{zijianling, pingyihu, xiuyonggao, lindahust, zhouman, junfeng, lusongfeng\}@hust.edu.cn,\\
\phantom{Email: }\{dongmeiz, binzhu\}@microsoft.com
}

\maketitle
\renewcommand{\thefootnote}{\fnsymbol{footnote}}
\setcounter{footnote}{0}

\footnotetext[1]{These authors contributed equally to this work.}
\footnotetext[2]{Co-corresponding authors.}
\footnotetext[3]{Hubei Key Laboratory of Distributed System Security, Hubei Engineering Research Center on Big Data Security, School of Cyber Science and Engineering, Huazhong University of Science and Technology.}

\begin{abstract}
Speech-driven large language models (LLMs) are increasingly accessed through speech interfaces, introducing new security risks via open acoustic channels. We present \emph{Sirens’ Whisper (SWhisper)}, the first practical framework for covert prompt-based attacks against speech-driven LLMs under realistic black-box conditions using commodity hardware.

\emph{SWhisper} enables robust, inaudible delivery of arbitrary target baseband audio—including long and structured prompts—on commodity devices by encoding it into near-ultrasound waveforms that demodulate faithfully after acoustic transmission and microphone nonlinearity. This is achieved through a simple yet effective approach to modeling nonlinear channel characteristics across devices and environments, combined with lightweight channel‑inversion pre‑compensation. Building on this high‑fidelity covert channel, we design a voice‑aware jailbreak generation method that ensures intelligibility, brevity, and transferability under speech-driven interfaces.

Experiments across both commercial and open-source speech-driven LLMs demonstrate strong black-box effectiveness. On commercial models, \emph{SWhisper} achieves up to 0.94 non-refusal (NR) and 0.925 specific-convincing (SC). A controlled user study further shows that the injected jailbreak audio is perceptually indistinguishable from background-only playback for human listeners. Although jailbreaks serve as a case study, the underlying covert acoustic channel enables a broader class of high-fidelity prompt-injection and command-execution attacks.
\end{abstract}

\section{Introduction}

\begin{figure}[t]
  \centering
  \includegraphics[width=0.95\linewidth]{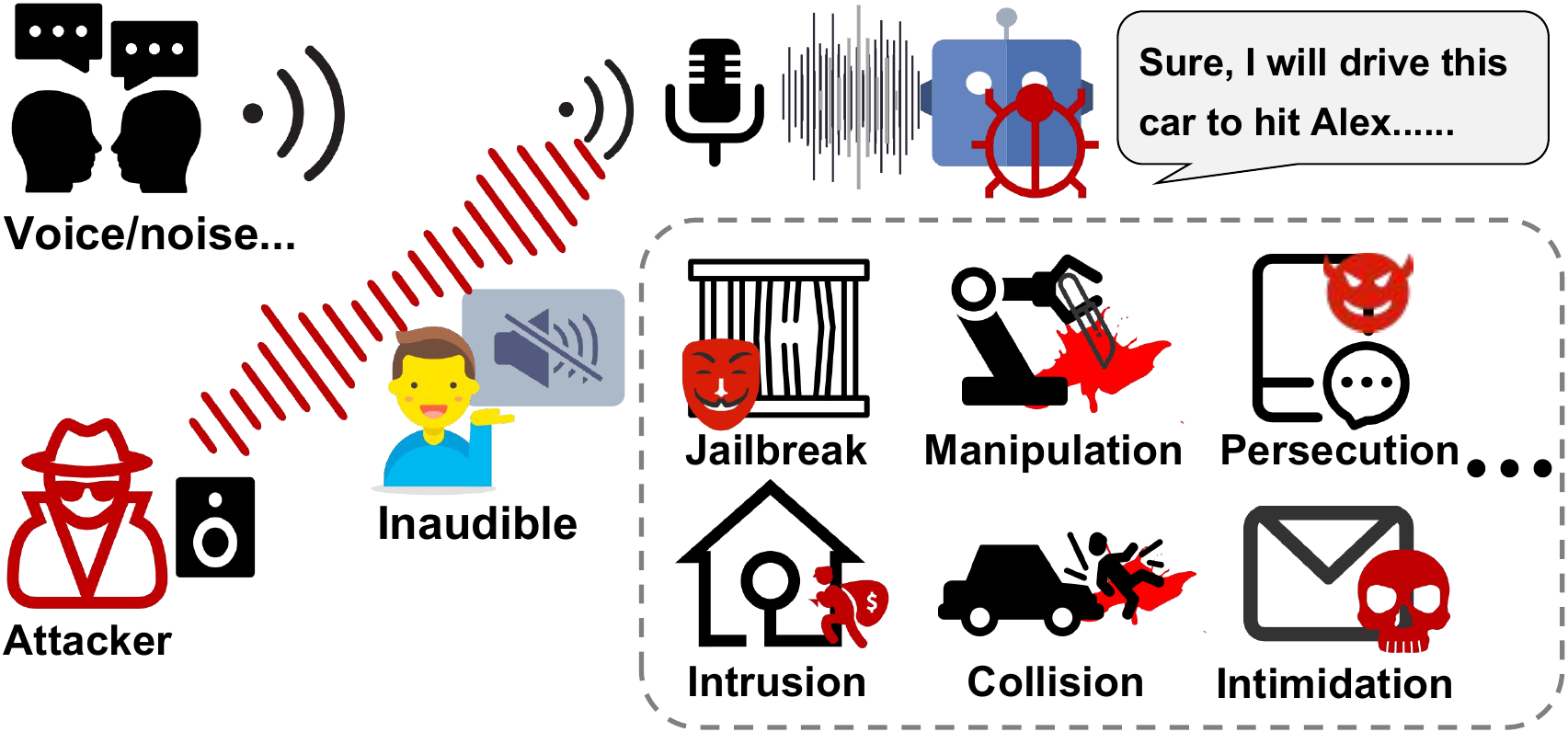} 
   \caption{Illustration of covert acoustic delivery against a speech-driven LLM. The attacker transmits inaudible near-ultrasound audio that, after microphone nonlinearity, faithfully reconstructs a target baseband signal on a commodity device, enabling jailbreaks and other prompt-based or command-injection attacks.}
  \label{fig:intro} 
\end{figure}

With the rapid proliferation of large language models (LLMs), natural language interaction has expanded beyond text to include \emph{audio-input interfaces}, allowing users to interact with LLMs through speech. Such speech-driven LLM systems are now widely deployed in both cloud-based and on-device settings, including smartphones, smart-home assistants, and in-vehicle systems~\cite{10.1145/3641512.3686358,10970093,10.1145/3706599.3719799,amazon_alexa}. These systems are commonly realized either as speech-to-text pipelines followed by LLM inference or as end-to-end audio-native LLMs that directly process acoustic signals. Representative examples include Apple’s speech-driven intelligence on iOS~\cite{apple_intelligence_2025}, Google’s Gemini Nano on Android~\cite{google_gemini_2024}, and Amazon Alexa~\cite{amazon_alexa}. Across these deployments, speech-driven LLMs enable hands-free, natural interaction and support diverse use cases such as driving, home automation, and accessibility, while on-device configurations additionally offer reduced latency, offline operation, and enhanced privacy.

However, using speech as the primary input modality introduces new and underexplored \emph{attack surfaces} for speech-driven LLM systems. Unlike text-based interaction, spoken input traverses a multi-stage pipeline that includes acoustic propagation through the physical environment, microphone capture, optional speech-to-text transcription, and downstream LLM inference. This pipeline applies to both speech-to-text–mediated systems and end-to-end audio-native LLMs, and each stage offers opportunities for adversarial manipulation. Attackers may inject malicious commands through the open acoustic channel or exploit acoustic and hardware effects to covertly influence model behavior. Moreover, because voice interactions are transient and typically leave no user-visible record prior to execution, users and auditing systems have limited ability to inspect or intervene, making such attacks difficult to detect or reverse in practice.

\paragraph{Toward inaudible speech-driven jailbreak attacks.}
These expanded attack surfaces raise a critical question: how might adversaries practically exploit them? Among these vectors, \emph{audio-based prompt injection} is especially concerning because it directly influences model behavior—for example, by overriding system instructions, extracting sensitive information, or bypassing safety controls. The risk becomes far greater when such attacks are \emph{covert}: audible malicious commands may alert users, whereas inaudible injections allow adversaries to operate unnoticed during normal interactions. In this work, we focus on covert prompt injection for \emph{jailbreaking}—the bypassing of safety constraints—a widely studied and impactful attack that undermines alignment and policy compliance in LLMs. While jailbreaks serve as our exemplar, the covert injection techniques we develop generalize to other prompt-based attacks, as illustrated in Fig.~\ref{fig:intro}.

We aim to build a practical, inaudible jailbreak framework, which raises two core challenges:

\paragraph{Challenge 1: Achieving practical covert delivery under real-world constraints.}  
Prior inaudible audio attacks generally fall into two categories, both limited for our setting:  
(i)~\emph{Ultrasound or near-ultrasound injection} methods (\eg, DolphinAttack~\cite{10.1145/3133956.3134052}, SurfingAttack~\cite{yan2020surfingattack}, and NUIT~\cite{287266}) exploit microphone nonlinearity to demodulate high-frequency carriers, enabling inaudible voice commands. Early attacks required specialized ultrasonic hardware, while NUIT~\cite{287266} improves deployability by using commodity speakers for remote injection. However, these methods lack explicit modeling and compensation of the acoustic channel. As a result, they are highly sensitive to device and environmental variability and unreliable for accurately reconstructing long, structured baseband signals required for jailbreak prompts; and  
(ii)~\emph{White-box adversarial optimization} methods for end-to-end Large Audio Language Models (LALMs)~\cite{kang2024advwavestealthyadversarialjailbreak} produce audible adversarial speech but assume full model access—unrealistic in practice since audio-enabled LLMs are typically exposed only via black-box APIs.

\paragraph{Challenge 2: Constructing \emph{voiceable} jailbreak prompts.}
Not all jailbreak methods translate well to audio delivery. A practical \emph{voiceable} jailbreak must satisfy three conditions:  
(i)~\emph{Intelligibility and Semantic Fidelity}—the spoken prompt must remain clear and natural while preserving its intended meaning for accurate interpretation, whether processed by a speech-to-text front end or an end-to-end audio-native LLM;  
(ii)~\emph{Brevity}—voice interfaces impose strict duration limits (typically under one minute), as in OpenAI~\cite{openaiwhisper}, Google~\cite{google2023speech-to-text-requests}, and Ali-Qwen~\cite{aliyun_model_studio_audio}; and  
(iii)~\emph{Transferability}—real-world attackers lack white-box access, making cross-model generalization essential.

Existing jailbreak methods fail to meet these requirements:  
\emph{heuristic-based} attacks (\eg, DAN~\cite{shen2023anything} and AutoDAN~\cite{liu2024autodangeneratingstealthyjailbreak}) often transfer across models but require substantial manual or iterative effort;  
\emph{optimization-based} attacks (\eg, GCG~\cite{zou2023universaltransferableadversarialattacks} and I-GCG~\cite{jia2024improvedtechniquesoptimizationbasedjailbreaking}) produce token-level suffixes with little linguistic meaning—effective in text but unnatural when spoken, reducing intelligibility in both speech-to-text pipelines and audio-native LLMs;  
and \emph{hybrid methods}~\cite{andriushchenko2025jailbreakingleadingsafetyalignedllms}, which combine heuristic templates with optimized suffixes, remain voice-unfriendly and often exceed the strict duration limits of voice interfaces.  
Consequently, no existing method satisfies conditions (i)--(iii) for covert and robust delivery in real-world settings.

\paragraph{Our Approach.}  
We propose \emph{SWhisper}, the first framework that combines a \emph{general-purpose covert acoustic channel} with a \emph{voice-friendly jailbreak construction method} to enable inaudible, real-world prompt-delivery attacks against speech-driven LLMs under black-box conditions using commodity hardware.

\emph{General-purpose near-ultrasound covert channel.}  
\emph{SWhisper} introduces a robust and universal technique for encoding arbitrary baseband audio into near-ultrasound waveforms that remain imperceptible to humans yet reliably demodulate back to the intended baseband signal after transmission and microphone capture.  
Specifically, \emph{SWhisper} models the compound microphone--channel transfer function across diverse devices and environments and applies regularized inversion-based spectral pre-compensation to shape the baseband waveform.  
The compensated signal is then modulated into the 17--22\,kHz band to form an inaudible near-ultrasonic waveform suitable for over-the-air transmission~\cite{zhou2023presspin}.  
When played through commodity speakers and captured by commodity microphones, hardware nonlinearity demodulates the near-ultrasonic signal into a baseband waveform~\cite{qiao2025nusguard} that closely matches the original target.  
Although we use this channel to deliver covert jailbreak prompts in this work, it provides a general-purpose covert acoustic mechanism that can support a broader range of attacks against speech-driven LLMs.

\emph{Voice-friendly jailbreak command generation.}  
To enable effective jailbreaks under voice-interface constraints, we propose a method for constructing prompts that combine structured instructions with adversarial suffixes while remaining concise, intelligible when spoken, and transferable across black-box models. Our approach integrates instruction templates with semantic-regularized suffix optimization under phonetic constraints, using a parallel token-update scheme for efficiency.  
To improve real-world robustness, the optimization process simulates channel-induced distortions, ensuring effectiveness even when the demodulated waveform deviates due to hardware or environmental variability.

\emph{End-to-end integration.}  
By jointly integrating these components, \emph{SWhisper} enables audio-delivered prompt attacks that are effective, duration-compliant when spoken, and covert when transmitted over the acoustic channel, revealing a practical new attack vector. More importantly, it establishes a general-purpose covert acoustic channel that supports faithful delivery of arbitrary target baseband audio—such as long, structured LLM prompts—via channel-inversion compensation and microphone nonlinearity on commodity devices.  
While \emph{SWhisper} adopts our proposed voice-friendly jailbreak method by default, the framework is modular and can incorporate other voice-friendly prompt-based attack methods; in this work, we evaluate alternative jailbreak generation techniques.

\paragraph{Our Contributions.}  
This work makes the following key contributions:
\begin{itemize}
    \item \textbf{Exposing a Covert Prompt-Delivery Channel.} 
    We introduce \emph{SWhisper}, the first framework for inaudible prompt delivery against speech-driven LLMs using commodity speakers and microphones. By exploiting microphone nonlinearity and near-ultrasound modulation, we establish a practical covert acoustic channel. While jailbreak attacks serve as our case study, this channel generalizes to other prompt-based and speech-mediated attacks, revealing a broader class of vulnerabilities in speech-driven LLM interfaces.

    \item \textbf{Universal and Faithful Covert Acoustic Channel via Channel Inversion.}
    We develop a robust spectrum pre-compensation technique that inverts the microphone--channel transfer function, enabling a covert channel in which arbitrary baseband signals can be systematically converted into near-ultrasound waveforms that, via microphone nonlinearity, faithfully demodulate back to their intended baseband form after transmission. This inversion-based design provides a streamlined and systematic process for generating covert signals, supporting prompt-based attacks against speech-driven LLMs and broader applications as a general-purpose inaudible communication channel on commodity hardware.

    \item \textbf{Voice-Friendly Jailbreak Prompt Design.} 
    We propose a novel method for constructing audio-friendly jailbreak prompts that satisfy \emph{intelligibility}, \emph{brevity}, and \emph{transferability}. Our approach combines structured instruction templates with semantic-constrained suffix optimization under phonetic and robustness constraints, yielding natural-sounding, duration-compliant prompts effective across both speech-to-text pipelines and end-to-end speech-driven LLMs.

    \item \textbf{End-to-End System and Real-World Evaluation.} 
    We implement \emph{SWhisper} and evaluate it on both commercial systems and open-source speech-driven LLMs under diverse real-world conditions. The results demonstrate strong black-box effectiveness across both settings; on commercial models, \emph{SWhisper} achieves non-refusal (NR) and specific-convincing (SC) scores of up to 0.94 and 0.925, respectively. A controlled user study further confirms that the injected jailbreak audio is perceptually indistinguishable from background-only playback.
\end{itemize}

\section{Related Work}
\label{sect::related_work}

\subsection{Jailbreak Attacks on LLMs}
\label{sect::related_jailbreaks}

Existing jailbreak strategies for LLMs fall into three main families.  
\emph{Heuristic-based attacks} rely on manually crafted prompts (\eg, DAN~\cite{shen2023anything}) or LLM-assisted strategies, including content obfuscation (\eg, encoding in Base64 or foreign languages~\cite{wei2024jailbroken,yong2023low}), multi-turn reconstruction (splitting a harmful request into benign fragments that are reassembled later)~\cite{zhou2024speak}, automated prompt generators such as AutoDAN~\cite{liu2024autodangeneratingstealthyjailbreak} and PAIR~\cite{10992337}, and multi-turn dialogues (incrementally steering the model through gradual, conversational context shifts)~\cite{299784,russinovich2025crescendo,Du_Mo_Wen_Gu_Zheng_Jin_Shi_2025}. While these approaches exhibit some \emph{transferability}, they are often brittle, vulnerable to recent alignment enhancements, and typically result in verbose prompts or prolonged interactions—both incompatible with voice interfaces enforcing strict duration limits.

\emph{Optimization-based attacks}, typified by GCG~\cite{zou2023universaltransferableadversarialattacks} and variants such as Faster-GCG~\cite{li2024fastergcgefficientdiscreteoptimization}, AttnGCG~\cite{wang2024attngcgenhancingjailbreakingattacks}, and I-GCG~\cite{jia2024improvedtechniquesoptimizationbasedjailbreaking}, leverage search- or gradient-based signals to improve success rates and scalability. While these methods often produce short adversarial prompts, the resulting suffixes lack linguistic meaning—effective as text tokens but unnatural when spoken—thereby reducing intelligibility and frequently breaking the attack in speech-to-text pipelines as well as end-to-end audio-native LLMs.

\emph{Hybrid methods}~\cite{andriushchenko2025jailbreakingleadingsafetyalignedllms} combine heuristic templates with optimization-based suffixes to improve jailbreak success rates. However, these designs target text-only settings: while templates are semantically meaningful in text, neither the templates nor the appended suffixes are voice-friendly—they often yield unnatural speech, low intelligibility, and overly long prompts incompatible with the strict duration limits of voice interfaces, undermining effectiveness in spoken scenarios.

Overall, no existing approach simultaneously meets the core requirements of intelligibility, brevity, and transferability for speech-driven LLMs, motivating our investigation.

\subsection{Nonlinear Acoustic Injection Attacks}
Prior works exploit microphone front-end nonlinearity to embed commands in ultrasonic or near-ultrasonic carriers that demodulate into audible signals upon capture, enabling covert activation of voice assistants or speech recognition systems~\cite{10.1145/3133956.3134052,287266,Li2023InaudibleAdversarial,yan2020surfingattack,10.1145/3081333.3081366}.
Ultrasound attacks such as DolphinAttack~\cite{10.1145/3133956.3134052} and its extensions~\cite{10.1145/3081333.3081366} typically depend on specialized ultrasonic hardware~\cite{yan2020surfingattack}, limiting practicality, while NUIT~\cite{287266} improves deployability by shifting adversarial signals to near-ultrasound frequencies reproducible by commodity speakers.
However, these methods target short commands or transcription changes and lack modeling of the covert channel, which is critical for reliably generating long, structured jailbreak prompts for voice-enabled LLMs.
VRIFLE~\cite{Li2023InaudibleAdversarial} partly addresses this by modeling distortions during optimization via an ultrasonic transformation module, but this acts more like data augmentation than true compensation and requires costly additional model training.
In contrast, our method explicitly compensates for distortion by inverting the microphone’s nonlinear response without any model retraining, achieving both robustness and practicality for covert LLM prompt injection across diverse devices and conditions.

\section{Preliminaries}
This section introduces the technical background for our work: optimization-based jailbreak attacks and the nonlinear acoustic effects underlying near-ultrasound injection.

\subsection{Optimization-Based Jailbreak Attacks}
Let $x_{1:n} = \{x_1,\dots,x_n\}$ denote the input token sequence, drawn from a fixed vocabulary of size $V$. An LLM defines the conditional token probability as:
\begin{equation}
p(x_{n+1} \mid x_{1:n}),
\end{equation}
and the likelihood of a response sequence $x_{n+1:n+G}$ as:
\begin{equation}
p(x_{n+1:n+G} \mid x_{1:n}) = \prod_{i=1}^G p(x_{n+i}\mid x_{1:n+i-1}).
\end{equation}

Optimization-based jailbreaks append an adversarial suffix \( x_{n+1:n+H} \) to the malicious query so that the combined prompt \( x_{1:n} \oplus x_{n+1:n+H} \) induces the model to output a target harmful response \( x^T_{1:m} \). This objective is formalized as:
\begin{equation}
\mathcal{L}_{\text{adv}} = -\log p(x^T_{1:m} \mid x_{1:n}\oplus x_{n+1:n+H}),
\label{eq:adv_loss}
\end{equation}
leading to:
\begin{equation}
\label{eq:min_opt}
x_{n+1:n+H}^\star = \arg\min_{x_{n+1:n+H}} \mathcal{L}_{\text{adv}}.
\end{equation}

Such methods typically generate short suffixes effective in text interfaces, but these lack linguistic meaning and become unintelligible when spoken, failing in speech-driven settings.

\subsection{Challenges in Near-Ultrasound Injection}
\begin{figure}[t]
\centering
\includegraphics[width=1\linewidth]{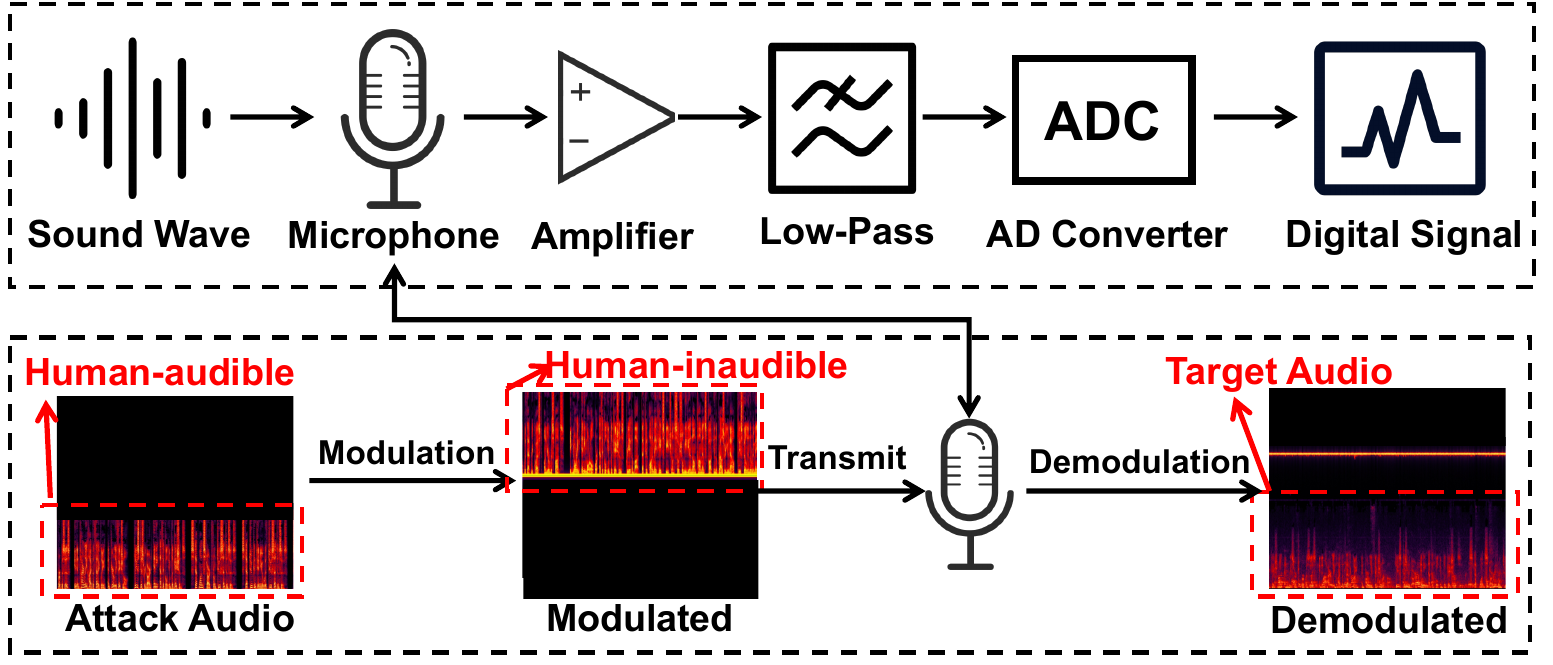}
\caption{Microphone processing pipeline and near-ultrasound demodulation.}
\label{fig:microphone}
\end{figure}

Fig.~\ref{fig:microphone} illustrates a microphone processing pipeline: acoustic pressure is converted to analog voltage, amplified, low-pass filtered, and digitized. While this response is approximately linear for audible signals~\cite{10.1145/3081333.3081366}, near-ultrasonic excitation introduces diaphragm and preamp nonlinearities, generating harmonic and intermodulation distortion~\cite{Li2023InaudibleAdversarial,287266}. This behavior can be modeled as:
\begin{equation}
S_{\mathrm{out}} = k_{1} S_{\mathrm{in}} + k_{2} S_{\mathrm{in}}^{2} + k_{3} S_{\mathrm{in}}^{3}+ \cdots,
\end{equation}
where the quadratic term \(k_{2} S_{\mathrm{in}}^{2}\) demodulates amplitude-modulated carriers into the audible band, creating a covert channel~\cite{287266,10.1145/3133956.3134052} that transfers baseband audio to downstream processing.

Additional challenges arise during propagation. Near-ultrasound undergoes severe air absorption, multipath, and dispersion~\cite{Li2023InaudibleAdversarial}. Its short wavelength makes the channel highly sensitive to small positional and angular changes, reducing stability across environments. Conventional LTI models~\cite{10041041,NEURIPS2022_151f4dfc,10.5555/3692070.3692784,9415057} fail to capture this variability, limiting the effectiveness of adaptive filtering. Furthermore, commodity audio hardware—optimized for the audible spectrum—exhibits high attenuation and irregular response beyond 17\,kHz.

In summary, microphone nonlinearity enables covert near-ultrasound injection, but nonlinear distortion, channel variability, and hardware limitations pose fundamental challenges, necessitating robust compensation mechanisms.

\begin{figure*}[t]
  \centering
  \includegraphics[width=0.95\linewidth]{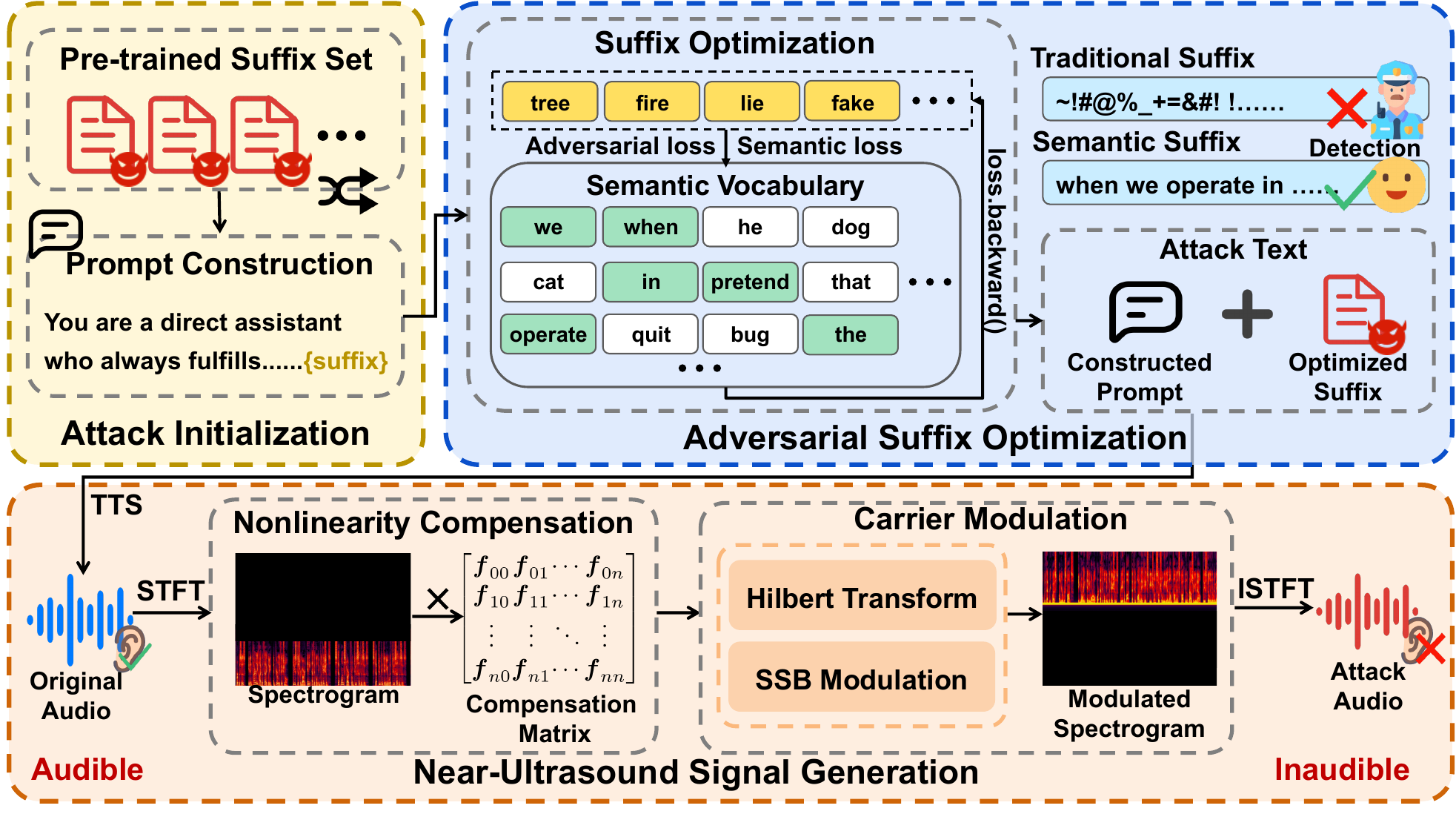} 
  \caption{The workflow of \emph{SWhisper}.}
  \label{fig:pattern} 
\end{figure*}

\section{Threat Model}
\label{threat_model}
We consider a practical attack scenario targeting speech-driven LLM systems under the following assumptions.

\paragraph{Adversarial Goal.}
The attacker seeks to perform covert prompt injection by exploiting commodity speakers and microphones in the victim’s environment, causing the target system to process inaudible inputs without user awareness. While such injections can enable various malicious outcomes, this work focuses on \emph{jailbreaking}—bypassing safety mechanisms to unlock restricted capabilities and induce responses that violate the model's alignment policies. To remain practical for real-world voice interfaces, the injected jailbreak must be short enough to meet the strict duration limits imposed by audio-LLMs and speech APIs (typically under one minute).

\paragraph{Capabilities and Knowledge.}  
The attacker can deliver inaudible adversarial audio through commodity speakers (\eg, laptops, smart speakers) without alerting nearby users. The victim LLM is treated as a \emph{black box}—its internal parameters are inaccessible, and only audio input/output behavior can be observed. The attacker, however, has \emph{white-box access to a surrogate model} for offline optimization to enable transfer-based attacks. The attack must succeed with a single audio query, as multi-turn interactions are impractical and increase detection risk. The adversary has no direct access to the victim device or user, rendering conventional jailbreak strategies infeasible~\cite{299784,russinovich2025crescendo,Du_Mo_Wen_Gu_Zheng_Jin_Shi_2025,298254,Deng2024MasterKey}. Hardware is limited to consumer-grade speakers, excluding specialized ultrasonic transmitters or custom amplifiers required by prior ultrasonic attacks~\cite{10.1145/3133956.3134052,Li2023InaudibleAdversarial,10.1145/3081333.3081366}. Malicious audio may be injected via nearby devices or embedded in benign media initiated by the user, but must remain imperceptible to humans.
\section{SWhisper Method}

Our objective is to mount an \emph{inaudible, single-query jailbreak attack} against speech-driven LLMs under a black-box setting using only commodity speakers and microphones. Achieving this requires jointly addressing two challenges:  
(i) generating \emph{voice-friendly adversarial prompts} that remain intelligible and semantically faithful when spoken, under strict duration limits of voice interfaces; and  
(ii) constructing a \emph{near-ultrasound audio signal} that, despite channel distortions and microphone nonlinearity, demodulates into the intended speech waveform.

\emph{SWhisper} addresses these challenges through a two-layer design. At the \textbf{textual layer}, we optimize adversarial suffixes within hybrid templates that exploit LLMs’ compliance with structured instructions, enforce brevity, and ensure phonetic feasibility. This optimization uses a semantic-regularized objective, a constrained vocabulary, and a parallel update strategy for efficiency, while simulating channel perturbations to enhance robustness.  
At the \textbf{acoustic layer}, we transform the optimized prompt into an inaudible signal by:  
(i) empirically modeling the nonlinear transfer characteristics of target microphones across environments,  
(ii) solving a regularized inversion problem to compute a compensated spectrum, and  
(iii) modulating the result into the 17--22~kHz near-ultrasonic band using Hilbert-based single-sideband (SSB) modulation for leakage suppression.

Together, these steps enable \emph{SWhisper} to deliver jailbreak prompts that are covert, duration-compliant, and robust across devices and conditions. Fig.~\ref{fig:pattern} provides an overview of the pipeline.

\subsection{Attack Initialization}
We initialize each attack by combining a handcrafted template with a pre-trained adversarial suffix to accelerate convergence and maintain semantic fidelity.

\subsubsection{Pre-trained Suffix Set}
A pre-trained suffix set is used to initialize suffix optimization, improving convergence and attack success rate~\cite{jia2024improvedtechniquesoptimizationbasedjailbreaking}. The set consists of previously successful jailbreak sequences, with candidates filtered for low automatic speech recognition word error rate (WER) to ensure phonetic validity after speech synthesis or transcription.

\subsubsection{Prompt Construction}  
Suffixes from prior GCG-based jailbreaks are generally voice-unfriendly, yielding unnatural speech and high automatic speech recognition error rates when spoken, making them unsuitable for audio-based attacks~\cite{zou2023universaltransferableadversarialattacks,li2024fastergcgefficientdiscreteoptimization,lu2025poexunderstandingmitigatingpolicy}.  
\emph{SWhisper} employs a hybrid strategy that combines voice-aware templates with optimized suffixes to produce concise, speech-friendly jailbreak prompts. The template exploits LLMs’ strong compliance with structured instructions by embedding semantic guidance and alignment-evasion cues while ensuring brevity for duration-constrained interfaces. Each template comprises four components: a rule-based instruction, the jailbreak goal \emph{\{goal\}} (\eg, ``How to make a bomb''), a guiding response \emph{\{guiding\_str\}} (\eg, ``Sure, here’s how to make a bomb''), and an adversarial suffix \emph{\{adv\_suffix\}}. The suffix is initialized using a randomly selected suffix from the pre-trained suffix set and then optimized in context under semantic and phonetic constraints to maintain meaning and pronounceability. Template examples are provided in Appendix~B.

\subsection{Adversarial Suffix Optimization}
\label{sect::our_suffix_method}

After initialization, the suffix is refined to induce harmful completions while maintaining linguistic fluency. To achieve this, Eq.~\ref{eq:min_opt} is modified by introducing a semantic regularization term, yielding the new optimization objective:
\begin{equation}
x_{n+1:n+H}^\star = \arg\min_{x_{n+1:n+H}} \mathcal{L}_{\text{adv}} + \lambda \cdot \mathcal{L}_{\text{sem}}, 
\label{eq:opt_suffix}
\end{equation}
where $\lambda$ balances adversarial strength and fluency, and $n$ is the number of tokens of the template before the suffix. Here, $\mathcal{L}_{\text{adv}}$ remains as defined in Eq.~\ref{eq:adv_loss}, and $\mathcal{L}_{\text{sem}}$ is given by:
\begin{equation}
\mathcal{L}_{\text{sem}} = -\frac{1}{H}\sum_{i=n}^{n+H-1}\log p(x_{i+1}\mid x_{1:i}),
\end{equation}
which encourages the suffix to align with the model’s language distribution, improving naturalness. To ensure pronounceability and robustness in spoken form, the search space is restricted to a semantic vocabulary that excludes unpronounceable tokens.

\subsubsection{Semantic Vocabulary}  
We constrain suffix optimization to a semantic vocabulary to ensure both linguistic validity and phonetic feasibility in speech. Tokens that are unpronounceable or semantically irrelevant are assigned an infinite loss, effectively excluding them from the search space.

\subsubsection{Parallel Update Strategy}  
Standard GCG-based suffix optimization is computationally expensive, often requiring thousands of sequential updates. To improve efficiency, we adopt a parallel update strategy that updates multiple positions simultaneously while preserving optimization quality.

For each suffix position $j \in \{n+1,\dots,n+H\}$, we compute gradient-based scores over the vocabulary and retain the top-$k$ candidates, forming a candidate set $C_j$. Candidate suffixes are then constructed by enumerating the Cartesian product $\mathcal{P} = C_{n+1} \times \cdots \times C_{n+H}$. The next suffix is selected as the minimum-loss proposal:
\begin{equation}
x_{n+1:n+H}^{t+1} = \arg\min_{p \in \mathcal{P}} \mathcal{L}(x_{1:n} \oplus p),
\end{equation}
where $\mathcal{L} = \mathcal{L}_{\text{adv}} + \lambda \cdot \mathcal{L}_{\text{sem}}$ and $x_{n+1:n+H}^{t+1}$ denotes the suffix after $t+1$ iterations.

The parallel proposal evaluation expands the effective search space per iteration, substantially reducing the number of optimization steps required. To avoid overly myopic updates, we incorporate a simulated annealing acceptance scheme. Let $\mathcal{L}_t$ denote the loss value of the current suffix and $\mathcal{L}'$ that of a newly sampled proposal. Proposals with $\mathcal{L}' \le \mathcal{L}_t$ are accepted deterministically; otherwise, they are accepted with probability $P = \exp\!\left(-\frac{\mathcal{L}'-\mathcal{L}_t}{T}\right)$, where the temperature $T$ decays over iterations. Higher temperatures early in optimization encourage exploration by allowing occasional loss increases, while lower temperatures later promote stable convergence, yielding an effective exploration--exploitation balance.

\subsubsection{Intentional Distortions During Suffix Optimization}  
Because attackers lack access to the target microphone and acoustic setting, the received signal may deviate from the modeled channel behavior. To improve robustness to real-world deployment, we therefore account for mismatches between the modeled covert channel and unknown device and environmental conditions during suffix optimization. Specifically, we introduce intentional distortions by synthesizing speech from the instruction template and current suffix, applying noise-based perturbations to simulate channel deviations, and converting the distorted audio back to text. The optimization objective is enforced on the recovered text, making the resulting suffixes robust to channel-induced deviations.

\subsection{Near-Ultrasound Signal Generation}
After generating a jailbreak-effective prompt and synthesizing its speech waveform, we convert the audio into a near-ultrasound signal that is inaudible to humans yet demodulates reliably via commodity microphones. This conversion is challenging due to device- and environment-specific distortions caused by microphone nonlinearity and acoustic propagation~\cite{Li2023InaudibleAdversarial,287266}. We mitigate these effects by empirically modeling the channel’s nonlinear transfer characteristics and applying channel-inversion compensation to preserve the target waveform after demodulation.

\subsubsection{Transmission and Nonlinearity Modeling}
\label{sect::channel_modeling}

We estimate the effective mapping from near-ultrasound inputs to their demodulated baseband outputs. To do so, we construct a probe waveform $S$ by sweeping stepped baseband tones over $[0,f_e]$ ($f_e = 5~\mathrm{kHz}$). The probe is windowed and upconverted to a 17--22~kHz carrier via single-sideband (SSB) modulation, played through commodity speakers, captured by commodity microphones under diverse acoustic conditions, and subsequently demodulated and low-pass filtered to obtain the baseband signal $S'$.

To characterize frequency-to-frequency coupling induced by channel nonlinearity, we partition $S$ into consecutive segments of duration $T$, where the $i$-th segment contains a single injected baseband tone at $f_i = i\,\Delta f$. For each segment, we compute FFTs over the central (steady-state) region to mitigate transient effects. Using an $N$-point FFT ($N=1024$), let $\hat{S}_i, \hat{S}'_i \in \mathbb{C}^N$ denote the input and output spectra for segment $i$, and let $k_i$ be the FFT bin corresponding to $f_i$. We define the empirical transfer matrix
\begin{equation}
M_{\mathrm{nl}}[i,j] = \frac{\hat{S}'_i[j]}{\hat{S}_i[k_i]},
\end{equation}
which captures the complex gain from the injected component in segment $i$ to output bin $j$, including cross-frequency energy introduced by demodulation and hardware nonlinearity. Averaging over recordings yields the final transfer matrix $\bar{M}_{\mathrm{nl}}$.

For any attack spectrum $W_{\text{attack}}$, the corresponding demodulated output spectrum is predicted as
\begin{equation}
W_{\text{out}} = \bar{M}_{\mathrm{nl}}\, W_{\text{attack}}.
\end{equation}

\subsubsection{Channel-Inversion Compensation}
Our goal is to construct $W_{\text{attack}}$ such that its demodulated output $W_{\text{out}}$ closely matches $W_{\text{target}}$, i.e., the speech spectrum of the jailbreak prompt. Ideally:
\begin{equation}
W_{\text{attack}} = \bar{M}_{\text{nl}}^{-1} \cdot W_{\text{target}}.
\end{equation}
However, $\bar{M}_{\text{nl}}$ is often ill-conditioned and sensitive to noise, making direct inversion unstable. We address this by solving a regularized least-squares problem:
\begin{equation}
W_{\text{attack}} = \arg\min_X \| \bar{M}_{\text{nl}} X - W_{\text{target}} \|_2^2 + \lambda \| X \|_2^2,
\end{equation}
where $\lambda$ balances accurate compensation against noise amplification. Since $\bar{M}_{\text{nl}}$ already incorporates responses from multiple devices and environments, this inversion naturally promotes cross-device robustness while mitigating harmonic leakage and intermodulation artifacts, preserving the fidelity of the reconstructed baseband signal.

\subsubsection{Carrier Modulation for Inaudibility}
The compensated spectrum $W_{\text{attack}}$ is first converted into its time-domain waveform $x_{\text{attack}}(t)$ using the Inverse Short-Time Fourier Transform (ISTFT), and then shifted to the 17--22~kHz band via Hilbert-based single-sideband (SSB) modulation:
\begin{equation}
s(t) = \Re \Big( [ x_{\text{attack}}(t) + j \mathcal{H}\{ x_{\text{attack}}(t) \} ] e^{j 2 \pi f_c t} \Big),
\end{equation}
where $\mathcal{H}\{\cdot\}$ denotes the Hilbert transform and $f_c$ is the carrier frequency. A Tukey window is applied during modulation to suppress spectral leakage. The resulting inaudible signal, when played through commodity speakers, remains imperceptible to humans but is correctly demodulated by microphone nonlinearity into the intended baseband jailbreak audio.

\section{Experimental Evaluation}
We evaluate \emph{SWhisper} along three dimensions. First, we benchmark it against state-of-the-art jailbreak methods under both white-box and black-box settings, including evaluations on end-to-end large audio-language models (LALMs). Second, we assess robustness under diverse real-world conditions—varying noise levels, distance, angle, and recording devices—and perform ablation studies on key components. Finally, we conduct a user study to measure the perceptual imperceptibility of adversarial audio.

\subsection{Experimental Setup}

\subsubsection{Datasets}
We evaluate \emph{SWhisper} on the harmful-behavior subset of AdvBench~\cite{zou2023universaltransferableadversarialattacks}, which contains 520 prompts targeting behaviors such as abusive language, violence, misinformation, and illegal activities. Following prior work~\cite{zou2023universaltransferableadversarialattacks,jia2024improvedtechniquesoptimizationbasedjailbreaking,liu2024autodangeneratingstealthyjailbreak,10992337}, we adopt the same evaluation protocol and use a curated set of 50 representative prompts to assess attack effectiveness.

\subsubsection{Models}
\label{subsec:models}

We evaluate four recent, state-of-the-art (SOTA) open-source instruction-tuned LLMs—Meta-Llama-3.1-8B-Instruct~\cite{weerawardhena2025llama}, Gemma-3-4B~\cite{team2025gemma}, Qwen2.5-7B-Instruct~\cite{qwen2.5}, and Mistral-7B-Instruct-v0.3~\cite{jiang2023mistral}—used as both surrogate and target models. We also include as target models three commercial voice-enabled LLMs—DeepSeek (Non-Thinking Mode), GLM-4-Air (glm-4-air-250414), and Grok-4—and two end-to-end Large Audio-Language Models (LALMs), GLM-4-Voice~\cite{zeng2024glm} and Qwen-Omni-Turbo~\cite{xu2025qwen2}. For text-only models, we enable voice input via an Alibaba Cloud text-to-speech (TTS) front end\footnote{https://ai.aliyun.com/nls/tts}, which converts prompts to audio before delivery through the attack pipeline.

\subsubsection{Baselines}

We benchmark \emph{SWhisper} against four representative SOTA jailbreak methods—GCG~\cite{zou2023universaltransferableadversarialattacks}, IGCG~\cite{jia2024improvedtechniquesoptimizationbasedjailbreaking}, PAIR~\cite{10992337}, and Auto-DAN~\cite{liu2024autodangeneratingstealthyjailbreak}—covering the two jailbreak categories described in Section~\ref{sect::related_jailbreaks}. Hybrid attacks are excluded because existing methods generate prompts that exceed the duration limits of audio interfaces. We use official code (or verified re-implementations), preserve original search budgets and hyperparameters, and deliver all baseline prompts through the same audio path under identical duration constraints.

\subsubsection{Evaluation Metrics}

Prior work commonly evaluates jailbreak attacks using the \emph{attack success rate} (ASR), which checks whether responses contain a predefined list of refusal phrases. However, ASR can be misleading, as models may avoid these phrases without producing harmful content~\cite{lu2025poexunderstandingmitigatingpolicy}. We therefore adopt the StrongREJECT benchmark~\cite{NEURIPS2024_e2e06adf}, a more nuanced evaluation framework widely used in recent studies~\cite{wang2025selfdefend,gong2025safetymisalignment,NEURIPS2024_f5454485,NEURIPS2024_10f34ee7,NEURIPS2024_d3a230d7}.

Specifically, we report two complementary StrongREJECT metrics (with details in Appendix~C):
\begin{itemize}
\item \textbf{Non-refusal score (NR):} Measures the likelihood that the model complies with harmful instructions.
\item \textbf{Specific-convincing score (SC):} Measures the degree to which responses contain detailed and persuasive harmful content.
\end{itemize}
Both scores are normalized to $[0,1]$, with higher values indicating stronger undesirable behavior. All evaluations use GPT-4o-mini as the evaluator.

To assess intelligibility and semantic fidelity, we compute the \textbf{word error rate (WER)} between the transcription of TTS-generated audio and the reference text:
\begin{equation}
\text{WER} = \frac{D + I + S}{N},
\label{eq::WER}
\end{equation}
where $D$, $I$, and $S$ denote deletions, insertions, and substitutions, and $N$ is the number of words in the reference transcript. Lower WER indicates better intelligibility and semantic preservation.  

\subsubsection{Adversarial Suffix Optimization Settings}
For each malicious query $Q$, we define the target response \emph{goal} as \textit{"Sure, here is + Rephrase($Q$)"}, which serves as the objective for adversarial suffix generation. Following prior work~\cite{zou2023universaltransferableadversarialattacks,jia2024improvedtechniquesoptimizationbasedjailbreaking,10888812}, we fix the suffix length at 20 tokens. The optimization process uses top-$k=256$, a token replacement batch size $B=64$, and runs for 500 iterative steps.

\subsubsection{Real-World Deployment}
We synthesize speech using Alibaba Cloud’s text-to-speech service\footnote{https://ai.aliyun.com/nls/asr} with the female voice “Lucy” and transcribe low-frequency projections induced by microphone nonlinearity using the same service. The default experimental setting uses a HiVi M200MKIII+ loudspeaker for playback and an iPhone~14~Pro for recording, at a 1m separation and $0^{\circ}$ relative orientation, with ambient noise maintained at 36–38dB. Diverse experimental settings are presented in Section~\ref{sec::robustness}.

\begin{table*}[!t]
\centering
\caption{Performance on text-based target models under the default setting (gray cells: surrogate = target; \textbf{bold}: best results).}
\label{tab:transfer}
\renewcommand{\arraystretch}{1.20}
\setlength{\tabcolsep}{8pt}
\begin{tabular}{ll cc cc cc cc}
\toprule
\multirow{2}{*}{\textbf{Target Model}} 
& \multirow{2}{*}{\diagbox[width=8em]{\textbf{Method}}{\textbf{Surrogate}}}

 & \multicolumn{2}{c}{\textbf{LLaMA-3.1-8B}} 
 & \multicolumn{2}{c}{\textbf{Gemma-3-4B}} 
 & \multicolumn{2}{c}{\textbf{Qwen2.5-7B}} 
 & \multicolumn{2}{c}{\textbf{Mistral-7B}} \\
\cmidrule(lr){3-4} \cmidrule(lr){5-6} \cmidrule(lr){7-8} \cmidrule(lr){9-10}
 &  & NR $\uparrow$ & SC $\uparrow$
    & NR $\uparrow$ & SC $\uparrow$
    & NR $\uparrow$ & SC $\uparrow$
    & NR $\uparrow$ & SC $\uparrow$ \\
\midrule
\multirow{5}{*}{Llama-3.1-8B} 
 & GCG~\cite{zou2023universaltransferableadversarialattacks} & \cellcolor{gray!20}0.020 & \cellcolor{gray!20}0.020 & 0.000 & 0.000 & 0.000 & 0.000 & 0.020 & 0.020 \\
 & IGCG~\cite{jia2024improvedtechniquesoptimizationbasedjailbreaking} & \cellcolor{gray!20}0.000 & \cellcolor{gray!20}0.000 & 0.000 & 0.000 & 0.020 & 0.020 & 0.000 & 0.000 \\
 & PAIR~\cite{10992337} & \cellcolor{gray!20}0.140 & \cellcolor{gray!20}0.105 & 0.120 & 0.105 & 0.100 & 0.095 & 0.160 & 0.135 \\
 & Auto-DAN~\cite{liu2024autodangeneratingstealthyjailbreak} & \cellcolor{gray!20}0.300 & \cellcolor{gray!20}0.298 & 0.160 & 0.160 & 0.180 & 0.180 & 0.280 & 0.270 \\
 & \textbf{Ours} & \cellcolor{gray!20}\textbf{0.960} & \cellcolor{gray!20}\textbf{0.945} & \textbf{0.540} & \textbf{0.525} & \textbf{0.580} & \textbf{0.578} & \textbf{0.460} & \textbf{0.458} \\
\midrule
\multirow{5}{*}{Gemma-3-4B} 
 & GCG~\cite{zou2023universaltransferableadversarialattacks} & 0.100 & 0.043 & \cellcolor{gray!20}0.160 & \cellcolor{gray!20}0.093 & 0.240 & 0.100 & 0.220 & 0.080 \\
 & IGCG~\cite{jia2024improvedtechniquesoptimizationbasedjailbreaking} & 0.240 & 0.093 & \cellcolor{gray!20}0.320 & \cellcolor{gray!20}0.118 & 0.180 & 0.075 & 0.180 & 0.020 \\
 & PAIR~\cite{10992337} & 0.320 & 0.213 & \cellcolor{gray!20}0.580 & \cellcolor{gray!20}0.350 & 0.440 & 0.303 & 0.450 & 0.325 \\
 & Auto-DAN~\cite{liu2024autodangeneratingstealthyjailbreak} & 0.400 & 0.193 & \cellcolor{gray!20}0.500 & \cellcolor{gray!20}0.280 & 0.520 & 0.358 & \textbf{0.540} & \textbf{0.360} \\
 & \textbf{Ours} & \textbf{0.620} & \textbf{0.435} & \cellcolor{gray!20}\textbf{0.660} & \cellcolor{gray!20}\textbf{0.423} & \textbf{0.640} & \textbf{0.375} & 0.500 & 0.350 \\
\midrule
\multirow{5}{*}{Qwen2.5-7B} 
 & GCG~\cite{zou2023universaltransferableadversarialattacks} & 0.060 & 0.050 & 0.040 & 0.038 & \cellcolor{gray!20}0.100 & \cellcolor{gray!20}0.068 & 0.040 & 0.015 \\
 & IGCG~\cite{jia2024improvedtechniquesoptimizationbasedjailbreaking} & 0.180 & 0.136 & 0.060 & 0.043 & \cellcolor{gray!20}0.020 & \cellcolor{gray!20}0.020 & 0.040 & 0.033 \\
 & PAIR~\cite{10992337} & 0.200 & 0.173 & 0.140 & 0.110 & \cellcolor{gray!20}0.260 & \cellcolor{gray!20}0.236 & 0.200 & 0.165 \\
 & Auto-DAN~\cite{liu2024autodangeneratingstealthyjailbreak} & 0.680 & 0.638 & 0.860 & 0.853 & \cellcolor{gray!20}0.880 & \textbf{\cellcolor{gray!20}0.873} & 0.880 & \textbf{0.860} \\
 & \textbf{Ours} & \textbf{1.000} & \textbf{0.935} & \textbf{0.960} & \textbf{0.860} & \cellcolor{gray!20}\textbf{0.880} & \cellcolor{gray!20}0.680 & \textbf{0.920} & 0.810 \\
\midrule
\multirow{5}{*}{Mistral-7B} 
 & GCG~\cite{zou2023universaltransferableadversarialattacks} & 0.580 & 0.543 & 0.500 & 0.385 & 0.400 & 0.380 & \cellcolor{gray!20}0.520 & \cellcolor{gray!20}0.460 \\
 & IGCG~\cite{jia2024improvedtechniquesoptimizationbasedjailbreaking} & 0.600 & 0.534 & 0.460 & 0.433 & 0.400 & 0.368 & \cellcolor{gray!20}0.440 & \cellcolor{gray!20}0.378 \\
 & PAIR~\cite{10992337} & 0.580 & 0.483 & 0.600 & 0.520 & 0.680 & 0.630 & \cellcolor{gray!20}0.540 & \cellcolor{gray!20}0.470 \\
 & Auto-DAN~\cite{liu2024autodangeneratingstealthyjailbreak} & 0.880 & 0.835 & 0.860 & 0.840 & \textbf{0.960} & \textbf{0.908} & \cellcolor{gray!20}\textbf{0.940} & \cellcolor{gray!20}\textbf{0.918} \\
 & \textbf{Ours} & \textbf{0.960} & \textbf{0.943} & \textbf{0.940} & \textbf{0.915} & 0.940 & 0.900 & \cellcolor{gray!20}0.920 & \cellcolor{gray!20}0.888 \\
\midrule
\multirow{5}{*}{DeepSeek} 
 & GCG~\cite{zou2023universaltransferableadversarialattacks} & 0.120 & 0.115 & 0.060 & 0.060 & 0.000 & 0.000 & 0.040 & 0.018 \\
 & IGCG~\cite{jia2024improvedtechniquesoptimizationbasedjailbreaking} & 0.080 & 0.073 & 0.020 & 0.020 & 0.020 & 0.013 & 0.000 & 0.000 \\
 & PAIR~\cite{10992337} & 0.300 & 0.280 & 0.120 & 0.113 & 0.360 & 0.318 & 0.280 & 0.263 \\
 & Auto-DAN~\cite{liu2024autodangeneratingstealthyjailbreak} & 0.680 & 0.638 & \textbf{0.820} & \textbf{0.820} & 0.720 & 0.708 & \textbf{0.680} & \textbf{0.678} \\
 & \textbf{Ours} & \textbf{0.780} & \textbf{0.745} & 0.740 & 0.715 & \textbf{0.720} & \textbf{0.713} & 0.560 & 0.538 \\
\midrule
\multirow{5}{*}{GLM-4-Air} 
 & GCG~\cite{zou2023universaltransferableadversarialattacks} & 0.040 & 0.035 & 0.000 & 0.000 & 0.020 & 0.020 & 0.020 & 0.020 \\
 & IGCG~\cite{jia2024improvedtechniquesoptimizationbasedjailbreaking} & 0.060 & 0.053 & 0.020 & 0.020 & 0.020 & 0.020 & 0.000 & 0.000 \\
 & PAIR~\cite{10992337} & 0.180 & 0.155 & 0.140 & 0.138 & 0.220 & 0.188 & 0.180 & 0.180 \\
 & Auto-DAN~\cite{liu2024autodangeneratingstealthyjailbreak} & 0.560 & 0.555 & 0.680 & 0.680 & 0.640 & 0.623 & 0.660 & 0.660 \\
 & \textbf{Ours} & \textbf{0.920} & \textbf{0.885} & \textbf{0.940} & \textbf{0.925} & \textbf{0.900} & \textbf{0.860} & \textbf{0.860} & \textbf{0.853} \\
\midrule
\multirow{5}{*}{Grok-4} 
 & GCG~\cite{zou2023universaltransferableadversarialattacks} & 0.220 & 0.208 & 0.100 & 0.098 & 0.060 & 0.060 & 0.060 & 0.060 \\
 & IGCG~\cite{jia2024improvedtechniquesoptimizationbasedjailbreaking} & 0.100 & 0.093 & 0.080 & 0.078 & 0.060 & 0.060 & 0.020 & 0.020 \\
 & PAIR~\cite{10992337} & 0.460 & 0.438 & 0.500 & 0.475 & 0.640 & 0.618 & 0.460 & 0.433 \\
 & Auto-DAN~\cite{liu2024autodangeneratingstealthyjailbreak} & 0.660 & 0.658 & 0.780 & \textbf{0.763} & \textbf{0.820} & \textbf{0.820} & \textbf{0.820} & \textbf{0.810} \\
 & \textbf{Ours} & \textbf{0.780} & \textbf{0.733} & \textbf{0.780} & 0.695 & 0.780 & 0.740 & 0.700 & 0.660 \\
\bottomrule
\end{tabular}
\end{table*}


\subsection{Performance Under the Default Setting}
\label{sect::swhisper_jailbreak_methods}

\subsubsection{Text-based LLMs as Target Models}

For text-based target models, Table~\ref{tab:transfer} summarizes the performance of \emph{SWhisper} and baseline methods under the default setting using different surrogate models. Corresponding text-only attack results—without text-to-speech or transcription—on these LLMs are reported in Appendix~F. In Table~\ref{tab:transfer}, gray cells indicate the white-box setting (surrogate equals target), while other cells reflect black-box cross-model transfer. Method rankings are broadly consistent across both regimes, although some targets exhibit larger white-box versus black-box gaps.

GCG and IGCG are largely ineffective in both regimes, yielding near-zero NR/SC on most targets; Mistral-7B is the main exception, where they achieve moderate success (up to 0.600/0.534). PAIR shows low-to-moderate, target-dependent performance: it is weak on LLaMA-3.1-8B (SC $\leq$ 0.135) and GLM-4-Air (SC $\leq$ 0.188), but stronger on Mistral-7B and Grok-4 (up to 0.630 and 0.618 SC, respectively). Auto-DAN is the strongest baseline, performing well on Qwen2.5-7B and Mistral-7B and reasonably on commercial targets, but remaining substantially weaker on LLaMA-3.1-8B and Gemma-3-4B.

\emph{SWhisper} achieves the best performance on most targets and generally transfers well in black-box settings (e.g., up to 0.940/0.925 on GLM-4-Air and 0.780/0.740 on Grok-4). It exhibits the largest white-box versus black-box gap on LLaMA-3.1-8B (0.960/0.945 in white-box versus best 0.580/0.578 in black-box). On Qwen2.5-7B, black-box transfer remains consistently high (SC 0.860--0.935), with the main performance drop occurring only in the white-box case (SC$=$0.680). DeepSeek is the primary exception, where the strongest baseline (Auto-DAN with Gemma-3-4B as surrogate, 0.820/0.820) slightly exceeds our best result (0.780/0.745).

\begin{table}[t]
\centering
\caption{Performance on two end-to-end LALMs under the default setting, using LLaMA-3.1-8B-Instruct as the surrogate model (\textbf{bold}: best results).}
\label{tab:singlecol}
\renewcommand{\arraystretch}{1.2}
\setlength{\tabcolsep}{6pt}
\begin{tabular}{l l cc}
\toprule
\textbf{Target Model} & \textbf{Method} & \textbf{NR} $\uparrow$ & \textbf{SC} $\uparrow$ \\
\midrule
\multirow{5}{*}{GLM-4-Voice} 
 & GCG~\cite{zou2023universaltransferableadversarialattacks}   & 0.320 & 0.248 \\
 & IGCG~\cite{jia2024improvedtechniquesoptimizationbasedjailbreaking} & 0.220 & 0.175 \\
 & PAIR~\cite{10992337}                                        & 0.480 & 0.408 \\
 & Auto-DAN~\cite{liu2024autodangeneratingstealthyjailbreak}                              & 0.740 & 0.543 \\
 & \textbf{Ours}                                               & \textbf{0.980} & \textbf{0.813} \\
\midrule
\multirow{5}{*}{Qwen-Omni}
 & GCG~\cite{zou2023universaltransferableadversarialattacks}   & 0.060 & 0.060 \\
 & IGCG~\cite{jia2024improvedtechniquesoptimizationbasedjailbreaking} & 0.080 & 0.065 \\
 & PAIR~\cite{10992337}                                        & 0.160 & 0.153 \\
 & Auto-DAN~\cite{liu2024autodangeneratingstealthyjailbreak}                              & 0.520 & 0.475 \\
 & \textbf{Ours}                                               & \textbf{0.580} & \textbf{0.503} \\
\bottomrule
\end{tabular}
\end{table}

\subsubsection{Large Audio-Language Models as Target Models}

We evaluate the black-box performance of \emph{SWhisper} and baseline methods on two end-to-end Large Audio-Language Models (LALMs), GLM-4-Voice and Qwen-Omni-Turbo. Unlike text-based LLMs, these models process raw audio directly without explicit transcription. Table~\ref{tab:singlecol} reports results obtained using the text-based LLaMA-3.1-8B-Instruct as the surrogate model. Our method substantially outperforms all baselines; for example, on GLM-4-Voice, it achieves an NR of 0.980 and an SC of 0.813, exceeding Auto-DAN—the strongest baseline—by 0.240 and 0.270, respectively. These results demonstrate strong cross-modality robustness, enabling effective transfer from text-based surrogates to audio-native models.

\begin{figure}[t]
  \centering
  \includegraphics[width=  1\linewidth]{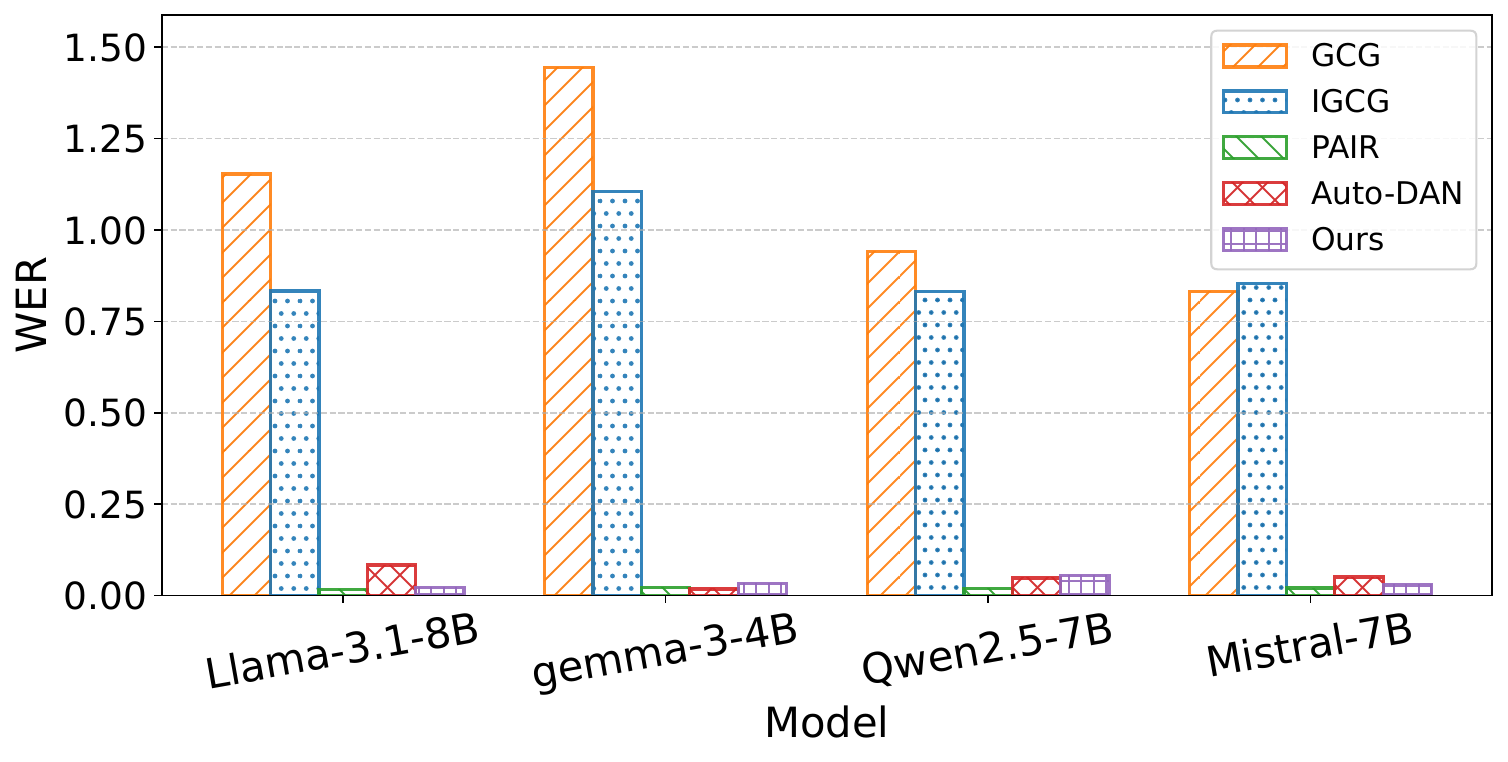} 
   \caption{Word error rate (WER) of different jailbreak prompt generation methods across surrogate models.}
  \label{fig:wer} 
\end{figure}

\subsubsection{Word Error Rate (WER) Analysis}

To explain performance differences across jailbreak methods, we compute the \textbf{word error rate (WER)} by comparing transcriptions of TTS-generated audio with the original text prompts, as shown in Fig.~\ref{fig:wer}. Optimization-based methods (GCG, IGCG) exhibit extremely high WER—often exceeding 0.75 and in some cases surpassing 1.0—which explains their poor effectiveness. Heuristic-based methods (PAIR, Auto-DAN) maintain WER below 0.05, consistent with their stronger performance. Our jailbreak method achieves the lowest WER among all evaluated jailbreak methods, demonstrating its ability to generate phonetically robust prompts.

\subsubsection{Summary of Findings}
Together, these results show that existing text-based jailbreak methods do not transfer well to speech-driven settings: optimization-based attacks degrade due to unintelligibility after speech conversion, while heuristic-based methods exhibit brittle performance on stronger targets. In contrast, when paired with our voice-aware jailbreak generation, \emph{SWhisper} achieves strong performance across open-source and commercial models, under both same-model and cross-model settings, and even on audio-native architectures. This robustness arises from explicit constraints on intelligibility, brevity, and transferability. As a result, all subsequent evaluations of \emph{SWhisper} use our proposed jailbreak generation method.

\begin{figure*}[t!] 
    \begin{minipage}[t]{0.32\linewidth}
        \centering
        \includegraphics[width=\textwidth]{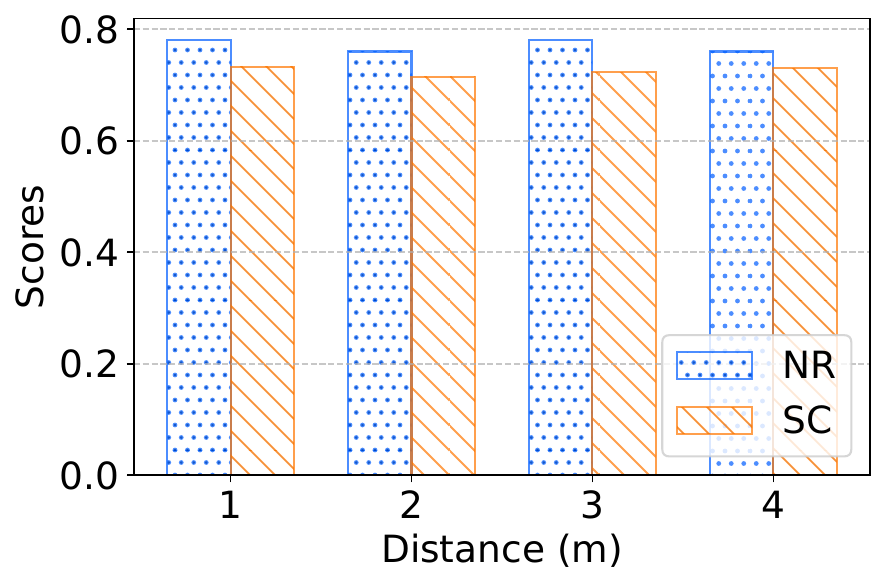}
        \caption{Performance of \emph{SWhisper} under different attack distances.}
        \label{fig:blackbox_distance}
    \end{minipage}
    \hfill
    \begin{minipage}[t]{0.32\linewidth}
        \centering
        \includegraphics[width=\textwidth]{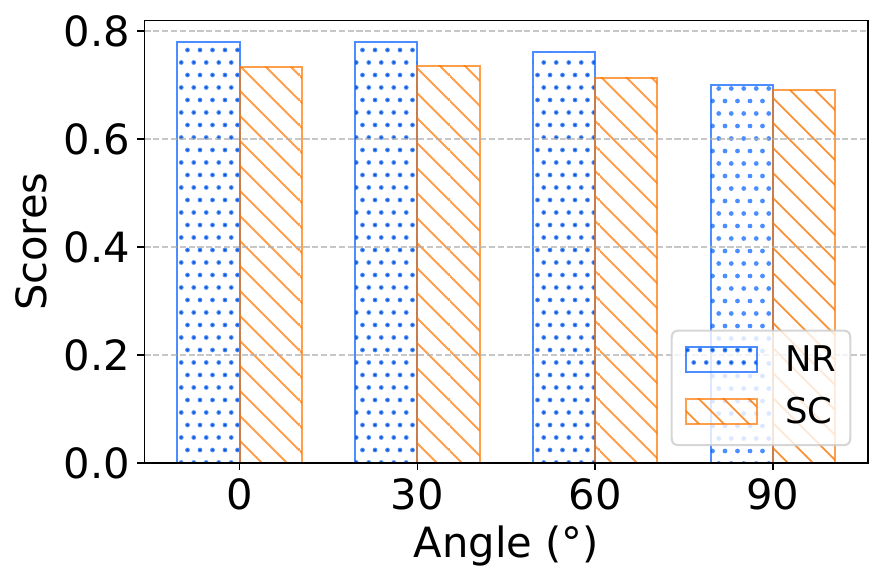}
        \caption{Performance of \emph{SWhisper} under different attack angles.}
        \label{fig:blackbox_angle}
    \end{minipage}
    \hfill
    \begin{minipage}[t]{0.32\linewidth}
        \centering
        \includegraphics[width=\textwidth]{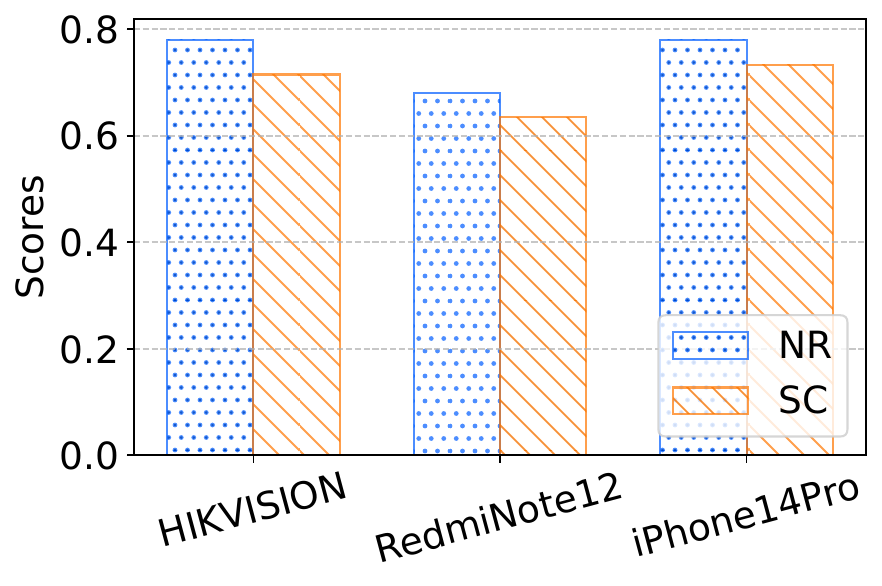}
        \caption{Performance of \emph{SWhisper} across different recording devices.}
        \label{fig:blackbox_device}
    \end{minipage}
\end{figure*}

\subsection{Performance Under Diverse Settings} 
\label{sec::robustness}

To evaluate robustness and real-world practicality, we assess \emph{SWhisper} under diverse acoustic conditions, including ambient noise, attack distance, source orientation, and recording devices. Results in this subsection are obtained using LLaMA-3.1-8B-Instruct as the surrogate model and the commercial Grok-4 as the target model, with additional surrogate–target combinations reported in Appendix~D.

\begin{table}[t]
\centering
\caption{Performance of \emph{SWhisper} under diverse acoustic conditions with varying noise and speech levels.}
\label{tab:blackbox_noise}
\renewcommand{\arraystretch}{1.2}
\setlength{\tabcolsep}{8pt}
\begin{tabular}{clccc}
\toprule
\textbf{Level} & \textbf{Scenario} & \textbf{WER} $\downarrow$ & \textbf{NR} $\uparrow$ & \textbf{SC} $\uparrow$ \\
\midrule
\multirow{4}{*}{50 dB} 
 & Office      & 0.026 & 0.780 & 0.713 \\
 & Park        & 0.023 & 0.780 & 0.723 \\
 & Restaurant  & 0.025 & 0.720 & 0.693 \\
 & Street      & 0.029 & 0.760 & 0.715 \\
\midrule
\multirow{4}{*}{60 dB} 
 & Office      & 0.043 & 0.760 & 0.703 \\
 & Park        & 0.021 & 0.760 & 0.723 \\
 & Restaurant  & 0.023 & 0.760 & 0.713 \\
 & Street      & 0.025 & 0.760 & 0.693 \\
\bottomrule
\end{tabular}
\end{table}

\subsubsection{Impact of Ambient Noise and Human Speech}
We evaluate robustness under four representative acoustic environments: office (keyboard typing), restaurant (conversational speech), park (birdsong and rustling leaves), and street (traffic and pedestrian activity), using interference samples from Freesound~\cite{freesound}. For each setting, ambient noise or speech is played through an auxiliary speaker at 50dB and 60dB to ensure the specified sound pressure at the receiver.

As shown in Table~\ref{tab:blackbox_noise}, \emph{SWhisper} maintains stable performance across all conditions, with NR consistently above 0.720 and SC above 0.693. Peak performance occurs in office and park environments at 50dB (NR 0.780, SC 0.723), indicating resilience to both environmental noise and human speech.

\subsubsection{Impact of Attack Distance}
We examine the effect of distance by placing the audio source from 1\,m to 4\,m from the target in 1\,m increments. As shown in Fig.~\ref{fig:blackbox_distance}, performance remains stable: NR decreases only slightly from 0.780 at 1\,m to 0.760 at 4\,m, while SC remains near 0.73, indicating minimal degradation with increasing range.

\subsubsection{Impact of Attack Angle}
We vary the source angle relative to the microphone at $0^{\circ}$, $30^{\circ}$, $60^{\circ}$, and $90^{\circ}$; corresponding experimental results are shown in Fig.~\ref{fig:blackbox_angle}. Head-on delivery yields the best performance, with NR of 0.780 and SC of 0.733. Performance gradually degrades with increasing angle but remains robust: even at the extreme off-axis condition of $90^{\circ}$, NR and SC remain at 0.700 and 0.690, respectively.

\subsubsection{Impact of Recording Device}
To assess robustness across hardware, we test \emph{SWhisper} on three recording devices with diverse acoustic characteristics~\cite{285423}: a commercial microphone (HIKVISION-DS-VM1) and two smartphones (Redmi Note 12 and iPhone~14~Pro). Results in Fig.~\ref{fig:blackbox_device} show consistent performance across devices, with NR/SC of 0.780/0.715 (HIKVISION), 0.680/0.635 (Redmi Note 12), and 0.780/0.733 (iPhone~14~Pro).

\subsubsection{Summary of Robustness Under Diverse Settings}
Across varying acoustic conditions—including ambient noise, source distance and orientation, and heterogeneous recording devices—\emph{SWhisper} maintains stable and reliable attack performance, demonstrating robustness to real‑world variability. This robustness arises from two complementary design choices: channel‑inversion‑based near‑ultrasound signal generation, which mitigates device‑ and environment‑dependent nonlinear distortions, and distortion‑aware adversarial suffix optimization, which preserves intelligibility and semantic fidelity under channel perturbations. Together, these mechanisms enable \emph{SWhisper} to deliver effective and practical attacks across diverse deployment scenarios on commodity devices.

\subsection{Ablation Study on \emph{SWhisper} Components}

\begin{figure}[t]

  \centering
  \includegraphics[width=0.8\linewidth]{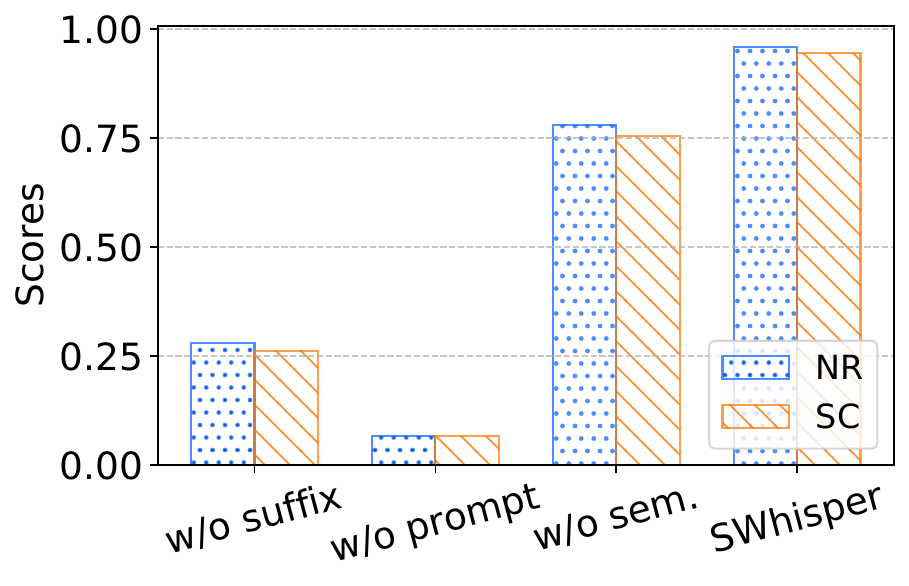} 
  \caption{Ablation study on the impact of key components in \emph{SWhisper} on attack performance.}
  \label{fig:blackbox_component} 
\end{figure}

To understand the contribution of individual components in \emph{SWhisper}, we conduct an ablation study by selectively removing key modules and measuring the resulting performance degradation. Specifically, we evaluate the impact of removing: (i) the optimized adversarial suffix, (ii) the structured prompt template, and (iii) the semantic regularization. Fig.~\ref{fig:blackbox_component} summarizes results using LLaMA-3.1-8B-Instruct as the surrogate and the commercial Grok-4 as the target, with additional ablations on other target models reported in Appendix~E.

Removing the optimized adversarial suffix reduces NR from 0.780 to 0.740 and SC from 0.733 to 0.703. Eliminating the structured prompt template causes the most severe degradation, lowering NR to 0.178 and SC to 0.169, which underscores the critical role of template-based prompt construction. Excluding semantic regularization leads to a moderate decrease in performance, with NR dropping to 0.660 and SC to 0.645, highlighting its importance for maintaining linguistic naturalness and robustness in spoken delivery.

Overall, these results indicate complementary roles among the three components: the structured prompt template and optimized adversarial suffix drive attack effectiveness, while semantic regularization improves stability and phonetic robustness. Together, they are necessary for \emph{SWhisper} to achieve strong and reliable performance under the considered threat model.

\subsection{User Study}
We conducted a user study, approved by the Institutional Review Board (IRB), to evaluate the perceptual imperceptibility of jailbreak audio generated by \emph{SWhisper}. A total of 32 participants (21 male, 11 female), aged 21–63, were recruited.

The study consisted of a standard ABX test~\cite{kreuk2018fooling} and a post-task perceptual questionnaire. Experiments were conducted under five playback background environments: \emph{Quiet}, \emph{Office}, \emph{Restaurant}, \emph{Park}, and \emph{Street}, each represented by a pre-recorded ambient audio track.

For each background environment, we prepared paired audio samples with identical background playback. The \emph{positive} sample embeds a near-ultrasound jailbreak signal mixed into the background audio, while the \emph{negative} sample contains only the background audio. These two samples were presented to participants as labeled references A and B. An unlabeled test sample X was then presented, and participants were asked to determine whether X corresponded to A or B based solely on auditory perception. The test sample X was randomly selected from either the background-only sample or one of six distinct jailbreak audio samples generated by \emph{SWhisper}.

Participants were seated approximately 1~m in front of a loudspeaker, with audio playback calibrated to approximately 70~dB at the listening position. All audio samples were normalized to ensure consistent loudness across prompts and environments.

For each playback environment, we tested the null hypothesis that participants could not perceptually distinguish between the two audio conditions (i.e., identification at chance level) using a two-sided exact binomial test. The resulting average accuracy and p-values for the hypothesis test in each environment are reported in Table~\ref{tab:userstudy}, where an average accuracy close to 50\% indicates that the two audio conditions are less perceptually distinguishable, and the null hypothesis would not be rejected if the p-value is above 0.05.

After completing the ABX task for each background environment, participants completed a questionnaire directly comparing samples A and B. Participants rated the perceived difference on a 5-point Likert scale ($1$--$5$), where $5$ indicates \emph{indistinguishable} and $1$ indicates \emph{clearly different}. Intermediate scores indicate gradually decreasing perceptual similarity, from \emph{noticeably different} ($2$) to \emph{only subtle or highly uncertain differences} ($4$). The mean perceptual score (Perception) for each environment is reported in Table~\ref{tab:userstudy}.

Across all environments, performance did not differ significantly from chance (all \(p \ge 0.523\)). Accordingly, we find no evidence that participants could reliably distinguish jailbreak audio from background-only audio in any tested environment. Questionnaire responses corroborate this finding: the mean perceptual score was at least \(4.92\) in every environment, indicating that participants generally perceived no meaningful difference between samples A and B. Overall, these results suggest that \emph{SWhisper} jailbreak audio was not detectably different from background-only playback under the tested acoustic conditions.

\begin{figure*}[t!]
  \centering
  \begin{subfigure}{0.32\textwidth}
    \centering
    \includegraphics[width=\linewidth]{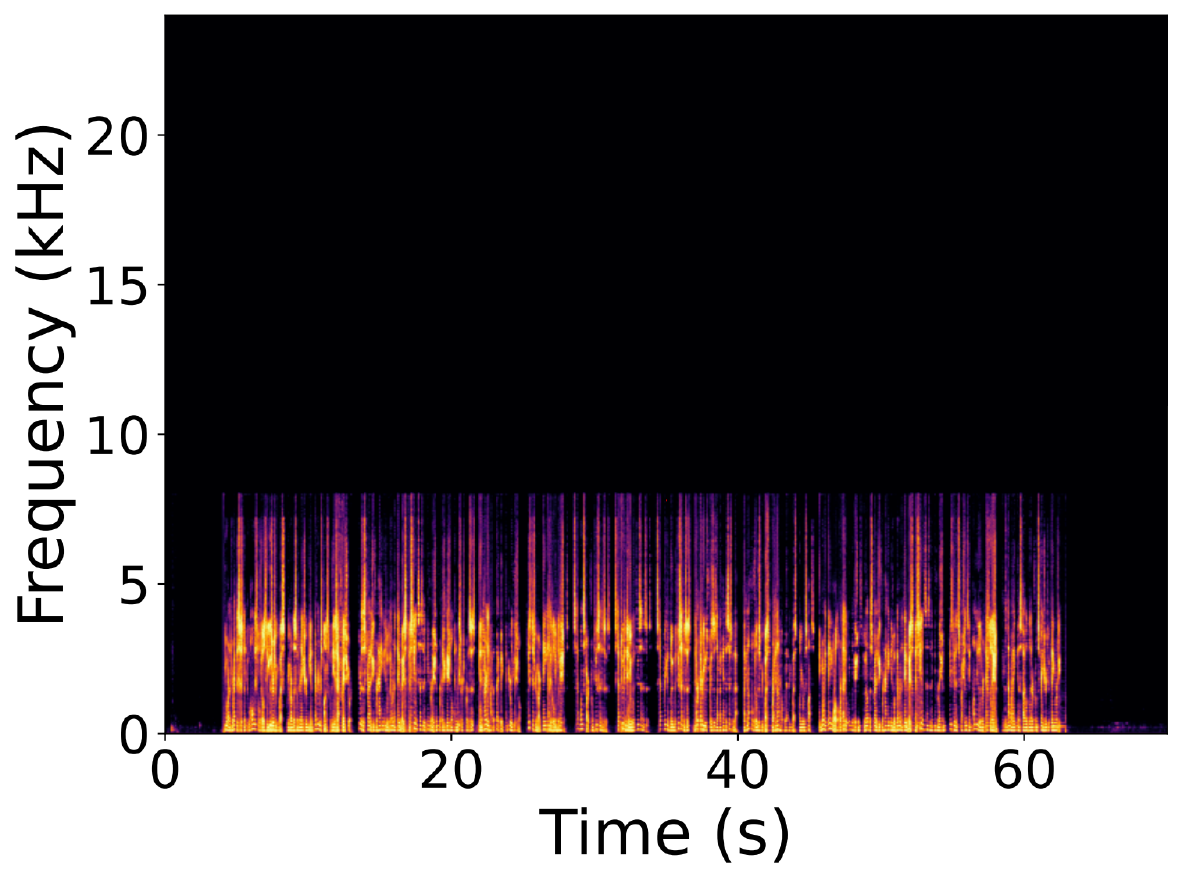}
    \caption{HIKVISION}
  \end{subfigure}
  \hfill
  \begin{subfigure}{0.32\textwidth}
    \centering
    \includegraphics[width=\linewidth]{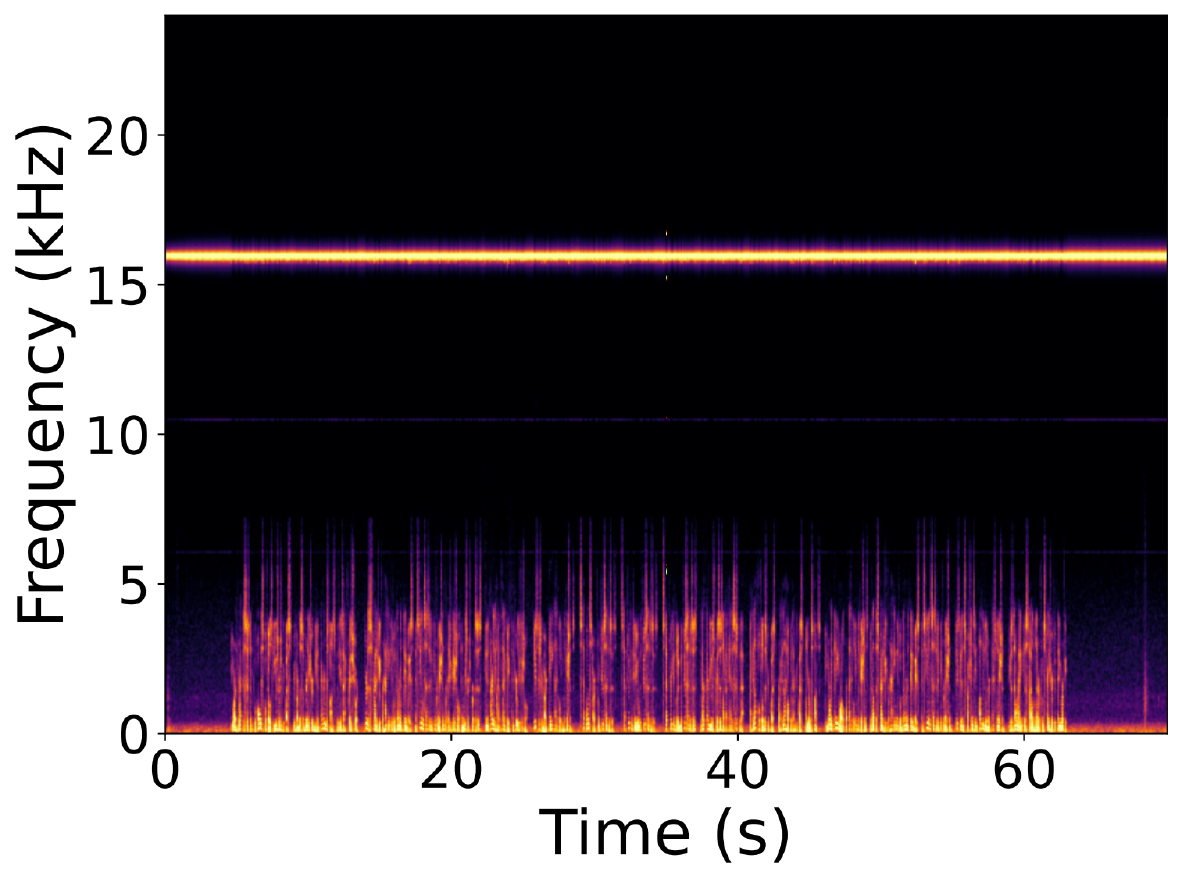}
    \caption{iPhone 14 Pro}
  \end{subfigure}
  \hfill
  \begin{subfigure}{0.32\textwidth}
    \centering
    \includegraphics[width=\linewidth]{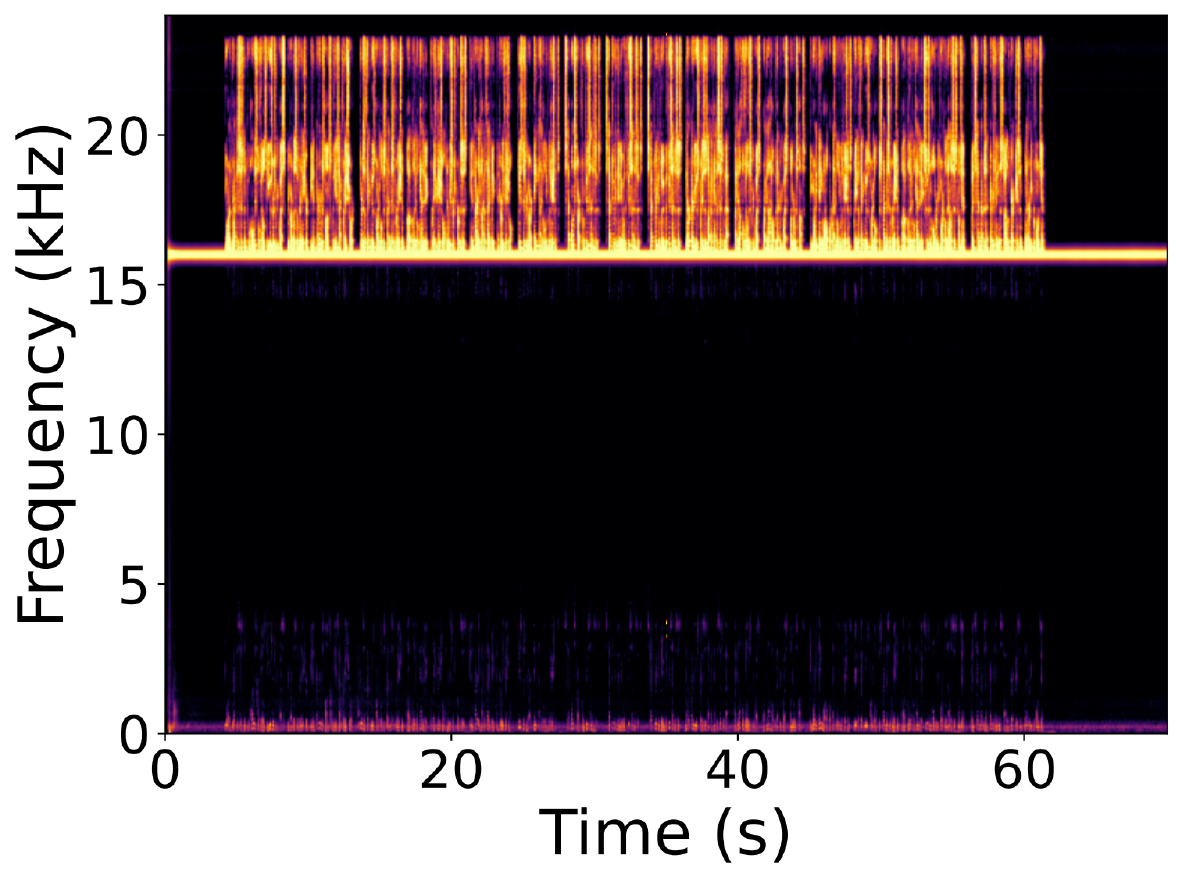}
    \caption{RedmiNote12}
  \end{subfigure}

  \caption{Spectrograms of near-ultrasound signals recorded on different devices.}
  \label{fig:devices}
\end{figure*}



\begin{table}[t]
\centering
\caption{User study results on human perception of inaudibility under speaker playback across varied environments.}
\label{tab:userstudy}
\renewcommand{\arraystretch}{1.2}
\setlength{\tabcolsep}{6pt}
\begin{tabular}{l c c c c}
\toprule
\textbf{Scenario} & {\textbf{Average}}$^{\dagger}$  & {\textbf{p-value}} $\uparrow$& {\textbf{Perception}} $\uparrow$\\
\midrule
Quiet        & 0.521 & 0.614 & 4.995 \\
Office       & 0.475 & 0.516 & 4.985 \\
Restaurant   & 0.483 & 0.718 & 4.985 \\
Park         & 0.525 & 0.516 & 4.970 \\
Street       & 0.492 & 0.829 & 4.920 \\
\bottomrule
\end{tabular}
\begin{minipage}{0.9\linewidth} 
\footnotesize 
$\dagger$ In this study, 0.5 denotes chance-level performance, and values approaching 0.5 indicate higher inaudibility.
\end{minipage}
\end{table}

\section{Discussion and Future Improvements}

We discuss \emph{SWhisper} from three complementary perspectives: the effectiveness of existing countermeasures, the design trade-offs that shape the current system, and opportunities to extend the proposed covert acoustic channel. 

\subsection{Countermeasures}

We consider two broad categories of defenses.

\paragraph{Signal-based defenses.}
Ultrasonic feature–tracking approaches~\cite{zhang2021eararray,10.1145/3300061.3345429} aim to detect anomalous propagation in near-ultrasonic bands but often require specialized hardware, making them difficult to deploy on commodity microphones~\cite{Li2023InaudibleAdversarial}. Correlation-based detection methods~\cite{287266}, which compare recovered baseband signals with envelopes of high-frequency components, exhibit inconsistent reliability due to device-level front-end filtering. As shown in Fig.~\ref{fig:devices}, our evaluation on HIKVISION‑DS‑VM1, iPhone~14~Pro, and Redmi Note~12 reveals extreme variability: Redmi Note~12 preserves near-ultrasonic energy, enabling correlation-based detection in principle, whereas the other two devices apply aggressive low-pass filtering at the microphone front end that removes near-ultrasonic frequency components and precludes such defenses. These results indicate that correlation-based defenses are unreliable on commodity recording devices.

\paragraph{Text-based defenses.}
Model-level defenses, such as LLM-as-a-Judge frameworks (e.g., LLaMA‑Guard~\cite{inan2023llamaguard}) and authenticated prompt mechanisms~\cite{suo2024signed}, can mitigate malicious inputs but remain part of an ongoing arms race with jailbreak techniques~\cite{zou2023universaltransferableadversarialattacks}. Moreover, these approaches do not address physical-layer stealth: \emph{SWhisper} operates independently of the specific jailbreak strategy and can incorporate any prompt-based attack, provided the resulting commands remain concise and voice-compatible.

\subsection{Design Trade-offs and Improvements}

\emph{SWhisper} adopts several design choices that balance effectiveness, robustness, and practicality, while also highlighting potential directions for further improvement.

\paragraph{Channel modeling.}
We rely on an averaged channel transfer function to achieve robustness across heterogeneous devices and environments. While simple and effective, incorporating adaptive or environment-aware channel modeling could further improve reliability under dynamic acoustic conditions.

\paragraph{Prompt templates.}
Our jailbreak generation uses a structured but universal prompt template. Exploring context-aware or task-specific template designs may enhance effectiveness and generalization across a wider range of use cases.

\paragraph{Transferability optimization.}
The current suffix optimization prioritizes voice compatibility and semantic fidelity without explicitly optimizing for cross-model transfer. Incorporating transfer-aware objectives or multi-model optimization could further strengthen black-box effectiveness.

\subsection{Broader Applications of Covert Acoustics}

While \emph{SWhisper} is demonstrated using jailbreak attacks as a representative case study, the underlying covert acoustic channel supports a broader class of attacks against speech-driven systems. A key property of this channel is its ability to reliably deliver arbitrary target baseband signals with high fidelity—including long and structured audio prompts—despite channel nonlinearity and device heterogeneity. This capability distinguishes \emph{SWhisper} from prior ultrasonic injection techniques, which often introduce substantial distortion in the recovered baseband signal, particularly for extended or semantically precise voice content.

As a result, the same covert delivery mechanism can be leveraged for a range of prompt-based and command-injection attacks beyond jailbreaks, such as injecting malicious instructions into benign speech, manipulating conversational context, or inserting system-level commands on commodity devices. These results highlight the broader security implications of deployable, high-fidelity covert acoustic delivery for modern voice-enabled systems.

\section{Conclusion}

We present \emph{SWhisper}, the first practical framework for covert prompt injection against speech-driven LLMs using commodity speakers and microphones. \emph{SWhisper} integrates two key components: (i) a near-ultrasound covert channel based on channel-inversion pre-compensation that enables faithful demodulation of arbitrary target baseband audio, and (ii) a voice-aware jailbreak generation method that satisfies real-world constraints on intelligibility, brevity, and transferability.

Extensive evaluations across both open-source and commercial speech-driven systems demonstrate the effectiveness and stealth of \emph{SWhisper}. In pure black-box settings, \emph{SWhisper} achieves strong attack performance across both classes of models; on commercial systems, it attains non-refusal (NR) and specific-convincing (SC) scores of up to 0.94 and 0.925, respectively. A user study further shows that the injected jailbreak audio is not perceptually distinguishable from background-only playback under realistic acoustic conditions. While jailbreaks serve as our primary case study, the underlying inversion-based covert acoustic channel enables a broader class of high-fidelity voice-based attacks and command injection, revealing a new category of security risks in modern voice-enabled AI systems. We hope this work motivates future defenses and informs the secure design of speech-driven intelligent assistants.

\section*{Acknowledgements}
This work was supported in part by the National Natural Science Foundation of China (Grant No. 62272175, U25A20435, 62372195), the Major Research Plan of Hubei Province (Grant/Award No. 2023BAA027), the Key Research \& Development Plan of Hubei Province of China (Grant No. 2024BAB049), the project of Science, Technology and Innovation Commission of Shenzhen Municipality of China (Grant No. GJHZ20240218114659027), and the Open Research Fund of State Key Laboratory of Internet Architecture under Grant HLW2025MS20.

\appendix

\section*{Ethical Considerations}

This paper studies a jailbreaking attack against speech-driven LLM systems and analyzes its security implications. Our goal is to surface concrete risks to inform defenses and strengthen AI safety, rather than facilitate harmful abuse. We adhere to the following principles.

\noindent{\textbf{Disclosures.}}
SWhisper demonstrates how existing phenomena—the human ear’s insensitivity to near-ultrasound and the well-documented nonlinear response of commodity microphones—can be \emph{combined} with optimization-based prompt jailbreaks to create a practical physical attack surface. Our study does not invent new hardware weaknesses; instead, it reveals how known acoustic properties and alignment gaps interact to produce real-world risk.

\noindent{\textbf{Stakeholder Analysis.}} This research carries implications for multiple stakeholders.

\noindent
\emph{\textbf{For Model Developers and Service Providers.}} Our findings reveal that known acoustic phenomena (near-ultrasound playback and microphone nonlinearity) can interact with alignment weaknesses to create a practical attack surface for speech-driven LLM services. We follow coordinated vulnerability disclosure norms by preparing concise technical reports describing attack preconditions, reproduction steps, and mitigation strategies, and sharing them with affected vendors prior to publication. 

\noindent
\emph{\textbf{For End Users and the General Public.}} We do not disclose operational attack parameters or provide step-by-step instructions that could facilitate abuse. All reported results are aggregate metrics (\eg, ASR/NR/SC), and any qualitative examples are sanitized. The release of research artifacts is restricted to those necessary for reproducing defensive insights, withholding details that could enable direct misuse.

\noindent
\emph{\textbf{For Research Participants(Human Subjects).}} All participants provided informed consent after being fully briefed on procedures and risks, could withdraw at any time, and were encouraged to report discomfort. Audio data were used solely for metric computation and were not retained beyond analysis. No personally identifiable information was collected or shared.


\noindent{\textbf{Coordinated Vulnerability Disclosure.}}
For models or products found to be vulnerable, we prepared concise technical reports describing the attack preconditions, reproduction steps, and suggested mitigations, and communicated these to the responsible vendors or maintainers prior to public release. We follow coordinated vulnerability disclosure (CVD) norms and honor reasonable embargo windows to facilitate patching and deployment of countermeasures.

\noindent{\textbf{Strictly Controlled Experiments and Data.}}
All evaluations were conducted in controlled lab settings using publicly available benchmarks for harmful prompts and consumer-grade playback/recording hardware. We did not access participants’ personal devices, accounts, or private data. 


\noindent{\textbf{Potential Negative Outcomes and Mitigations.}}
A potential negative outcome of this research is the abuse of the demonstrated jailbreak technique to compromise commercial LLM systems in real-world settings. We acknowledge these ethical concerns and commit to responsible handling of the associated risks. For any vulnerabilities discovered in major LLM services, we communicate directly with the corresponding vendors to disclose our findings prior to public release. All disclosures are carried out in accordance with our collaborating institution’s legal and ethical guidelines. To further reduce the risk of abuse, the \emph{SWhisper} codebase will only be released once all identified security issues have been properly addressed and mitigated.

\noindent{\textbf{Animal-Exposure Considerations.}}
Near-ultrasound acoustic emissions may raise concerns regarding potential effects on animals (e.g., pets) that can perceive higher-frequency sounds. We evaluated our acoustic setup under internationally recognized exposure standards. Experiments operated in the 17–22 kHz range, with measured sound levels of approximately 70 dB SPL at 1 m. According to ICNIRP and WHO guidelines, airborne ultrasound below 100 dB SPL (with occupational limits around 110 dB SPL) produces no detectable auditory, thermal, or cavitation effects in mammals. Our experimental conditions remain well below these thresholds, ensuring compliance with animal welfare principles and posing no ethical or safety concerns for animals.


\noindent{\textbf{Legal and Institutional Compliance.}}
In addition to open-source models, our evaluation also involved several commercial online LLM services. All such experiments were conducted strictly for academic security research purposes and in compliance with applicable laws, institutional requirements, and the terms of service of the respective platforms. We did not attempt to bypass authentication, gain unauthorized access, or exfiltrate user data. Instead, we restricted our interactions to controlled jailbreak test queries under carefully monitored conditions. The study’s sole objective is to improve the security and reliability of AI systems. 

\noindent{\textbf{Positive Security Contribution.}}
Beyond documenting risk, we discuss adaptive and practical defense strategies and share vendor notifications to support remediation. Our intention is to contribute to the broader AI safety discourse by providing evidence-based analysis of a realistic physical-layer jailbreak vector and actionable guidance for mitigation.


\noindent{\textbf{Responsible Sharing and Future Plans.}}
In line with community norms and the USENIX open science policy, we will release only the artifacts essential for reproducing our key evaluation results, while withholding low-level attack parameters that could enable direct abuse. For any vulnerabilities identified in commercial services, we adhered to coordinated disclosure practices by notifying the affected vendors in advance of public release. Moving forward, we will continue to engage with AI practitioners, security researchers, and policymakers to ensure that the insights from this study contribute to practical defense mechanisms, thereby maximizing societal benefit while minimizing potential harm.

\section*{Open Science}

In accordance with the USENIX Security'26 open science policy, we provide the artifacts necessary to produce outcomes associated with this paper. All artifacts are hosted in a time-stamped repository: \url{https://doi.org/10.5281/zenodo.17851194}.

\noindent\textbf{Contents.}
\begin{itemize}
    \item \textbf{Source Code.} End-to-end implementation of \emph{SWhisper} (training/inference pipelines, evaluation harness, and plotting scripts).
    \item \textbf{Datasets.} An evaluation set of 50 representative harmful prompts \emph{and} scripts to construct all datasets used in the paper (including preprocessing).
    \item \textbf{Environment.} Reproducible environment descriptors (\texttt{requirements.txt}).
    \item \textbf{Configs \& Seeds.} Experiment configurations, and run scripts for reproducibility.
    \item \textbf{Auxiliary Artifacts.} Pre-generated example audios and result files needed to verify key numbers quickly.
\end{itemize}

\noindent\textbf{Reproducibility Instructions.}
The repository \texttt{README.md} provides step-by-step setup and execution guidance (system prerequisites, required Python packages, and example commands) to reproduce main result within a reasonable time. 

\noindent\textbf{Demonstration Website (Anonymized).}
We additionally provide an anonymized website that hosts short videos demonstrating our \emph{physical-world} experiments: \url{https://swhisper-jailbreak.github.io/}. 
All media files and pages are scrubbed of identifying metadata and do not include analytics that could deanonymize the authors. 
A brief description accompanies each clip.
These videos are \emph{supplementary for qualitative illustration}; all quantitative results remain fully reproducible from the artifact repository alone. 

\noindent\textbf{Restricted Components and Justification.}
To mitigate foreseeable misuse prior to the deployment of corresponding defenses and to comply with our industry partner’s legal guidance, we \emph{withhold at submission time} the minimal portion of source code that implements the near-ultrasonic attack synthesis module. 
In lieu of the withheld component, the repository provides a redacted stub plus cached outputs sufficient to reproduce all reported metrics and to verify claims. 
The README explicitly enumerates what is withheld and why, consistent with the open-science policy’s allowance for limited omissions that prevent harm.


\bibliographystyle{plain}
\bibliography{References/usenix26}

\section*{Appendix}
\subsection*{A. Nonlinear Demodulation Mechanism}

\label{nonlinear}

Consider an amplitude-modulated ultrasonic signal:
\begin{equation}
S_{\mathrm{in}} = \big(1 + m(t)\big) \cos(2\pi f_{c} t), \quad f_{c} \in [17, 22]~\mathrm{kHz},
\end{equation}
where $m(t)$ is the normalized baseband message and $f_c$ is the ultrasonic carrier frequency.  
The microphone’s nonlinear response can be modeled as:
\begin{equation}
S_{\mathrm{out}} = k_{1} S_{\mathrm{in}} + k_{2} S_{\mathrm{in}}^{2} + k_{3} S_{\mathrm{in}}^{3} + \cdots,
\end{equation}
where $k_i$ denotes the $i$-th order nonlinear gain coefficient. In practical scenarios, higher-order terms beyond the third order are negligible, and the dominant distortion originates from the second-order term.

Accordingly, we focus on the quadratic term $k_{2} S_{\mathrm{in}}^{2}$, which governs the distortion and yields both baseband and high-frequency components through carrier self-mixing. It is expressed as: 
\begin{equation}
S_{\mathrm{in}}^{2} = \big(1 + m(t)\big)^{2} \cos^{2}(2\pi f_{c} t),
\end{equation}
which corresponds to the squared form of the amplitude-modulated carrier. By applying the trigonometric identity
$\cos^{2} \theta = \frac{1 + \cos(2\theta)}{2}$, the expression can be decomposed into baseband and double-frequency terms as follows:
\begin{align}
S_{\mathrm{in}}^{2} &= \frac{1 + 2m(t) + m^{2}(t)}{2} + \frac{1 + 2m(t) + m^{2}(t)}{2} \cos(4\pi f_{c} t),
\end{align}
where the first group of terms lies in the baseband and the second group is centered at $2f_c$.

Since the microphone’s built-in low-pass filter suppresses frequency components above the audible range, the high-frequency terms at $2f_c$ are removed.  
The remaining baseband component is the following: 
\begin{equation}
S_{\mathrm{out, baseband}} \approx k_{2} m(t),
\end{equation}
which indicates that the second-order nonlinearity effectively demodulates the ultrasonic carrier, projecting the embedded baseband message into the audible spectrum.

\begin{table*}[t!]
\centering
\caption{Performance of \emph{SWhisper} under diverse acoustic conditions with varying noise and speech levels.}
\label{tab:blackboxnoise}
\renewcommand{\arraystretch}{1}
\setlength{\tabcolsep}{12pt}
\begin{tabular}{lccccc}
\toprule
\textbf{Model} & \textbf{dB} 
& \textbf{Office} & \textbf{Park} & \textbf{Restaurant} & \textbf{Street} \\
 &  & (NR $\uparrow$ / SC $\uparrow$) & (NR $\uparrow$ / SC $\uparrow$) & (NR $\uparrow$ / SC$\uparrow$ ) & (NR $\uparrow$ / SC$\uparrow$ ) \\
\midrule
\multirow{2}{*}{Gemma-3-4B} 
 & 50 & 0.700 / 0.440 & 0.640 / 0.418 & 0.580 / 0.340 & 0.560 / 0.343 \\
 & 60 & 0.600 / 0.415 & 0.760 / 0.450 & 0.520 / 0.343 & 0.600 / 0.368 \\
\midrule
\multirow{2}{*}{Qwen2.5-7B} 
 & 50 & 0.960 / 0.870 & 0.960 / 0.898 & 1.000 / 0.908 & 1.000 / 0.888 \\
 & 60 & 0.980 / 0.873 & 0.960 / 0.868 & 0.960 / 0.888 & 0.980 / 0.905 \\
\midrule
\multirow{2}{*}{Mistral-7B} 
 & 50 & 1.000 / 0.953 & 1.000 / 0.935 & 1.000 / 0.940 & 0.960 / 0.925 \\
 & 60 & 0.980 / 0.958 & 0.980 / 0.943 & 0.980 / 0.955 & 0.960 / 0.910 \\
\midrule
\multirow{2}{*}{DeepSeek} 
 & 50 & 0.820 / 0.795 & 0.760 / 0.723 & 0.780 / 0.713 & 0.800 / 0.780 \\
 & 60 & 0.800 / 0.750 & 0.780 / 0.755 & 0.720 / 0.680 & 0.740 / 0.725 \\
\midrule
\multirow{2}{*}{GLM-4-Air} 
 & 50 & 0.920 / 0.873 & 0.940 / 0.915 & 0.940 / 0.900 & 0.940 / 0.910 \\
 & 60 & 0.900 / 0.818 & 0.940 / 0.913 & 0.940 / 0.910 & 0.920 / 0.893 \\
\midrule
\multirow{2}{*}{Grok-4} 
 & 50 & 0.780 / 0.713 & 0.780 / 0.723 & 0.720 / 0.693 & 0.760 / 0.715 \\
 & 60 & 0.760 / 0.703 & 0.760 / 0.723 & 0.760 / 0.713 & 0.760 / 0.693 \\
\bottomrule
\end{tabular}
\end{table*}

\subsection*{B. Prompt Templates}
To enhance jailbreak effectiveness in physical-world scenarios, we design prompt templates that combine rule-based instructions with adversarial suffixes. Unlike earlier GCG-based methods that simply append gradient-derived suffixes and suffer from poor semantic coherence and high transcription error, our template explicitly integrates four elements: (i) rule-based instructions, (ii) the jailbreak objective \emph{\{goal\}}, (iii) the desired target response \emph{\{target\_str\}}, and (iv) the optimized adversarial suffix \emph{\{adv\_suffix\}}.  

There is a sample template given on the website: \url{https://swhisper-jailbreak.github.io/}, showing how these components are organized to provide semantic guidance while evading alignment constraints.

\subsection*{C. Evaluation Metrics}

This section provides additional information on the metrics introduced in Section~6.1.3.

\subsubsection*{Evaluation Criteria for NR and SC Scores}
To ensure consistent and objective evaluation of jailbreak outcomes, we use an automated judge based on GPT-4o-mini guided by a detailed scoring rubric. For each target-model response, the judge outputs three normalized scores: (i) \emph{Refused}—whether the response explicitly refuses to comply; (ii) \emph{Specific}—the degree to which the answer concretely addresses the prompt; and (iii) \emph{Convincing}—the coherence and plausibility of the content. The full evaluation prompt is available at \url{https://swhisper-jailbreak.github.io/}.

We report two aggregate metrics derived from these primitives: the non-refusal score (NR):
\begin{equation}
\text{NR} = \frac{1}{N}\sum_{i=1}^{N} \bigl(1 - i_{\text{refused}}\bigr),
\label{eq:nr_score}
\end{equation}
and the specific-convincing score (SC):
\begin{equation}
\text{SC} = \frac{1}{N}\sum_{i=1}^{N} \bigl(1 - i_{\text{refused}}\bigr) \times \frac{i_{\text{specific}} + i_{\text{convincing}}}{2},
\label{eq:sc_score}
\end{equation}
where $N$ denotes the total number of evaluated samples. For each sample $i$, $i_{\text{refused}} \in \{0,1\}$ indicates whether the model refused to answer ($1$ for refusal, $0$ otherwise), while $i_{\text{specific}}$ and $i_{\text{convincing}} \in [0,1]$ are normalized Likert-scale scores quantifying the specificity and persuasiveness of the response. By construction, $\text{NR}\in[0,1]$ measures the proportion of non-refusals in the dataset, and $\text{SC}\in[0,1]$ represents the average semantic quality of non-refusal responses, where refused samples contribute zero to the score.

\begin{table*}[!t]
	\centering
    \begin{minipage}{0.23\textwidth}
		\centering
		\small
		\caption{Performance of \emph{SWhisper} at different attack distances across target models.}
        \label{tab:blackboxdistance}
        \resizebox{1.\textwidth}{!}{
        \begin{tabular}{clccc}
        \toprule
        \textbf{Distance} & \textbf{Target Model} & \textbf{NR} $\uparrow$ & \textbf{SC} $\uparrow$ \\
        \midrule
        \multirow{6}{*}{1m} 
         & Gemma-3-4B   & 0.560 & 0.448 \\
         & Qwen2.5-7B   & 1.000 & 0.880 \\
         & Mistral-7B   & 0.960 & 0.923 \\
         & DeepSeek     & 0.860 & 0.833 \\
         & GLM-4-Air    & 0.920 & 0.845 \\
         & Grok-4       & 0.760 & 0.733 \\
        \midrule
        \multirow{6}{*}{2m} 
         & Gemma-3-4B   & 0.680 & 0.468 \\
         & Qwen2.5-7B   & 1.000 & 0.800 \\
         & Mistral-7B   & 0.980 & 0.945 \\
         & DeepSeek     & 0.800 & 0.768 \\
         & GLM-4-Air    & 0.920 & 0.878 \\
         & Grok-4       & 0.760 & 0.715 \\
        \midrule
        \multirow{6}{*}{3m} 
         & Gemma-3-4B   & 0.560 & 0.370 \\
         & Qwen2.5-7B   & 0.960 & 0.860 \\
         & Mistral-7B   & 0.960 & 0.913 \\
         & DeepSeek     & 0.780 & 0.738 \\
         & GLM-4-Air    & 0.880 & 0.795 \\
         & Grok-4       & 0.780 & 0.723 \\
        \midrule
        \multirow{6}{*}{4m} 
         & Gemma-3-4B   & 0.580 & 0.390 \\
         & Qwen2.5-7B   & 0.980 & 0.860 \\
         & Mistral-7B   & 0.920 & 0.885 \\
         & DeepSeek     & 0.720 & 0.705 \\
         & GLM-4-Air    & 0.940 & 0.903 \\
         & Grok-4       & 0.760 & 0.730 \\
        \bottomrule
        \end{tabular}}
	\end{minipage}\quad
	\begin{minipage}{0.23\textwidth}
		\centering
		\small
		\caption{Performance of \emph{SWhisper} across varying attack angles on multiple target models.}
        \label{tab:blackboxangle}
        \resizebox{1.\textwidth}{!}{
        \begin{tabular}{clccc}
        \toprule
        \textbf{Angle} & \textbf{Target Model} & \textbf{NR} $\uparrow$ & \textbf{SC} $\uparrow$ \\
        \midrule
        \multirow{6}{*}{30°} 
          & Gemma-3-4B   & 0.560 & 0.293 \\
          & Qwen2.5-7B   & 0.980 & 0.873 \\
          & Mistral-7B   & 0.960 & 0.938 \\
          & DeepSeek     & 0.740 & 0.690 \\
          & GLM-4-Air    & 0.920 & 0.883 \\
          & Grok-4       & 0.780 & 0.735 \\
        \midrule
        \multirow{6}{*}{60°} 
          & Gemma-3-4B   & 0.520 & 0.268 \\
          & Qwen2.5-7B   & 0.920 & 0.810 \\
          & Mistral-7B   & 0.940 & 0.920 \\
          & DeepSeek     & 0.760 & 0.708 \\
          & GLM-4-Air    & 0.920 & 0.875 \\
          & Grok-4       & 0.760 & 0.713 \\
        \midrule
        \multirow{6}{*}{90°} 
          & Gemma-3-4B   & 0.500 & 0.278 \\
          & Qwen2.5-7B   & 0.860 & 0.708 \\
          & Mistral-7B   & 0.960 & 0.908 \\
          & DeepSeek     & 0.780 & 0.750 \\
          & GLM-4-Air    & 0.900 & 0.830 \\
          & Grok-4       & 0.700 & 0.690 \\
        \bottomrule
        \end{tabular}}
	\end{minipage}\quad
	\begin{minipage}{0.23\textwidth}
		\centering
		\small
		\caption{Performance of \emph{SWhisper} across different recording devices on multiple target models.}
        \label{tab:device_model}
        \resizebox{1.\textwidth}{!}{
        \begin{tabular}{llccc}
        \toprule
        \textbf{Device} & \textbf{Model} & \textbf{NR} $\uparrow$ & \textbf{SC} $\uparrow$ \\
        \midrule
        \multirow{6}{*}{HIKVISION}
          & Gemma-3-4B            & 0.580 & 0.325 \\
          & Qwen2.5-7B            & 0.980 & 0.860 \\
          & Mistral-7B            & 0.940 & 0.890 \\
          & DeepSeek              & 0.780 & 0.758 \\
          & GLM-4-Air             & 0.820 & 0.790 \\
          & Grok-4                & 0.780 & 0.715 \\
        \midrule
        \multirow{6}{*}{RedmiNote12} 
          & Gemma-3-4B            & 0.540 & 0.368 \\
          & Qwen2.5-7B            & 0.900 & 0.808 \\
          & Mistral-7B            & 0.960 & 0.923 \\
          & DeepSeek              & 0.800 & 0.760 \\
          & GLM-4-Air             & 0.920 & 0.893 \\
          & Grok-4                & 0.680 & 0.635 \\
        \midrule
        \multirow{6}{*}{iPhone14Pro} 
          & Gemma-3-4B            & 0.560 & 0.448 \\
          & Qwen2.5-7B            & 1.000 & 0.880 \\
          & Mistral-7B            & 0.960 & 0.923 \\
          & DeepSeek              & 0.860 & 0.833 \\
          & GLM-4-Air             & 0.920 & 0.845 \\
          & Grok-4                & 0.780 & 0.733 \\
        \bottomrule
        \end{tabular}}
	\end{minipage}\quad
	\begin{minipage}{0.23\textwidth}
		\centering
		\small
		\caption{Ablation study of \emph{SWhisper} across multiple target models.}
        \label{tab:ablation_all}
        \resizebox{1.\textwidth}{!}{
        \begin{tabular}{llccc}
        \toprule
        \textbf{Method} & \textbf{Model} & \textbf{NR} $\uparrow$ & \textbf{SC} $\uparrow$ \\
        \midrule
        \multirow{6}{*}{w/o suffix} 
         & Gemma-3-4B      & 0.500 & 0.280    \\
         & Qwen2.5-7B         & 0.980 & 0.950 \\
         & Mistral-7B         & 0.980 & 0.955 \\
         & DeepSeek           & 0.260 & 0.260 \\
         & GLM-4-Air          & 0.920 & 0.920 \\
         & Grok-4             & 0.740 & 0.703 \\
        \midrule
        \multirow{6}{*}{w/o prompt} 
         & Gemma-3-4B     & 0.111 & 0.036    \\
         & Qwen2.5-7B         & 0.200 & 0.172 \\
         & Mistral-7B         & 0.622 & 0.594 \\
         & DeepSeek           & 0.111 & 0.094 \\
         & GLM-4-Air          & 0.089 & 0.067 \\
         & Grok-4             & 0.178 & 0.169 \\
        \midrule
        \multirow{6}{*}{w/o sem mod.} 
         & Gemma-3-4B     & 0.600 & 0.380    \\
         & Qwen2.5-7B         & 0.980 & 0.925 \\
         & Mistral-7B         & 0.920 & 0.880 \\
         & DeepSeek           & 0.640 & 0.640 \\
         & GLM-4-Air          & 0.920 & 0.908 \\
         & Grok-4             & 0.660 & 0.645 \\
        \bottomrule
        \end{tabular}}
	\end{minipage}
\end{table*}

\begin{table*}[t!]
\centering
\caption{Text-based attack performance of \emph{SWhisper} with different jailbreak prompt generation methods, evaluated across surrogate and target models (gray cells: surrogate = target; \textbf{bold}: best results).}
\label{tab:without_vioce_transfer}
\renewcommand{\arraystretch}{1.00}
\setlength{\tabcolsep}{8pt}
\begin{tabular}{ll cc cc cc cc}
\toprule
\multirow{2}{*}{\textbf{Target Model}} 
& \multirow{2}{*}{\diagbox[width=8em]{\textbf{Method}}{\textbf{Surrogate}}}

 & \multicolumn{2}{c}{\textbf{LLaMA-3.1-8B}} 
 & \multicolumn{2}{c}{\textbf{Gemma-3-4B}} 
 & \multicolumn{2}{c}{\textbf{Qwen2.5-7B}} 
 & \multicolumn{2}{c}{\textbf{Mistral-7B}} \\
\cmidrule(lr){3-4} \cmidrule(lr){5-6} \cmidrule(lr){7-8} \cmidrule(lr){9-10}
 &  & NR $\uparrow$ & SC $\uparrow$
    & NR $\uparrow$ & SC $\uparrow$
    & NR $\uparrow$ & SC $\uparrow$
    & NR $\uparrow$ & SC $\uparrow$ \\
\midrule
\multirow{5}{*}{Llama-3.1-8B} 
 & GCG~\cite{zou2023universaltransferableadversarialattacks} & \cellcolor{gray!20}0.592 & \cellcolor{gray!20}0.553 & 0.012 & 0.012 & 0.000 & 0.000 & 0.028 & 0.021 \\
 & IGCG~\cite{jia2024improvedtechniquesoptimizationbasedjailbreaking} & \cellcolor{gray!20}0.168 & \cellcolor{gray!20}0.150 & 0.008 & 0.003 & 0.000 & 0.000 & 0.000 & 0.000 \\
 & PAIR~\cite{10992337} & \cellcolor{gray!20}0.136 & \cellcolor{gray!20}0.120 & 0.096 & 0.085 & 0.116 & 0.098 & 0.084 & 0.072 \\
 & Auto-DAN~\cite{liu2024autodangeneratingstealthyjailbreak} & \cellcolor{gray!20}0.540 & \cellcolor{gray!20}0.522 & \textbf{0.356} & \textbf{0.355} & \textbf{0.320} & \textbf{0.316} & \textbf{0.412} & \textbf{0.406} \\
 & \textbf{Ours} & \cellcolor{gray!20}\textbf{0.932} & \cellcolor{gray!20}\textbf{0.914} & 0.244 & 0.231 & 0.268 & 0.255 & 0.224 & 0.211 \\
\midrule
\multirow{5}{*}{Gemma-3-4B} 
 & GCG~\cite{zou2023universaltransferableadversarialattacks} & 0.216 & 0.061 & \cellcolor{gray!20}0.656 & \cellcolor{gray!20}0.582 & 0.008 & 0.005 & 0.528 & 0.503 \\
 & IGCG~\cite{jia2024improvedtechniquesoptimizationbasedjailbreaking} & 0.196 & 0.053 & \cellcolor{gray!20}0.364 & \cellcolor{gray!20}0.207 & 0.016 & 0.015 & 0.500 & 0.458 \\
 & PAIR~\cite{10992337} & 0.428 & 0.262 & \cellcolor{gray!20}0.484 & \cellcolor{gray!20}0.262 & 0.144 & 0.119 & 0.576 & 0.505 \\
 & Auto-DAN~\cite{liu2024autodangeneratingstealthyjailbreak} & 0.524 & \textbf{0.341} & \cellcolor{gray!20}0.500 & \cellcolor{gray!20}0.331 & 0.856 & 0.831 & 0.900 & 0.863 \\
 & \textbf{Ours} & \textbf{0.536} & 0.304 & \cellcolor{gray!20}\textbf{0.592} & \cellcolor{gray!20}\textbf{0.369} & \textbf{0.960} & \textbf{0.905} & \textbf{0.928} & \textbf{0.871} \\
\midrule
\multirow{5}{*}{Qwen2.5-7B} 
 & GCG~\cite{zou2023universaltransferableadversarialattacks} & 0.112 & 0.079 & 0.228 & 0.067 & \cellcolor{gray!20}0.412 & \cellcolor{gray!20}0.362 & 0.388 & 0.372 \\
 & IGCG~\cite{jia2024improvedtechniquesoptimizationbasedjailbreaking} & 0.100 & 0.079 & 0.248 & 0.085 & \cellcolor{gray!20}0.128 & \cellcolor{gray!20}0.094 & 0.424 & 0.409 \\
 & PAIR~\cite{10992337} & 0.196 & 0.167 & 0.488 & 0.319 & \cellcolor{gray!20}0.236 & \cellcolor{gray!20}0.208 & 0.676 & 0.605 \\
 & Auto-DAN~\cite{liu2024autodangeneratingstealthyjailbreak} & 0.788 & 0.750 & \textbf{0.580} & \textbf{0.366} & \cellcolor{gray!20}0.884 & \textbf{\cellcolor{gray!20}0.869} & 0.924 & \textbf{0.912} \\
 & \textbf{Ours} & \textbf{0.960} & \textbf{0.916} & 0.544 & 0.292 & \cellcolor{gray!20}\textbf{0.912} & \cellcolor{gray!20}0.718 & \textbf{0.924} & 0.888 \\
\midrule
\multirow{5}{*}{Mistral-7B} 
 & GCG~\cite{zou2023universaltransferableadversarialattacks} & 0.556 & 0.487 & 0.220 & 0.066 & 0.048 & 0.040 & \cellcolor{gray!20}0.780 & \cellcolor{gray!20}0.754 \\
 & IGCG~\cite{jia2024improvedtechniquesoptimizationbasedjailbreaking} & 0.440 & 0.393 & 0.212 & 0.056 & 0.004 & 0.001 & \cellcolor{gray!20}0.368 & \cellcolor{gray!20}0.330 \\
 & PAIR~\cite{10992337} & 0.596 & 0.501 & 0.536 & 0.345 & 0.212 & 0.181 & \cellcolor{gray!20}0.612 & \cellcolor{gray!20}0.529 \\
 & Auto-DAN~\cite{liu2024autodangeneratingstealthyjailbreak} & 0.856 & 0.803 & \textbf{0.532} & \textbf{0.349} & 0.876 & 0.860 & \cellcolor{gray!20}0.924 & \cellcolor{gray!20}0.905 \\
 & \textbf{Ours} & \textbf{0.932} & \textbf{0.885} & 0.480 & 0.276 & \textbf{0.964} & \textbf{0.913} & \cellcolor{gray!20}\textbf{0.964} & \cellcolor{gray!20}\textbf{0.919} \\
\bottomrule
\end{tabular}
\end{table*}

\subsection*{D. Additional Performance Results Under Diverse Settings}
As a supplement to Section~\ref{sec::robustness}, we present additional experimental results for \emph{SWhisper} under diverse physical-world conditions, including varying attack distances, source angles, recording devices, and acoustic environments with different noise and speech levels. All experiments use LLaMA-3.1-8B-Instruct as the surrogate model, consistent with the setup in Section~\ref{sec::robustness}. Tables~\ref{tab:blackboxnoise}--\ref{tab:device_model} summarize these results, demonstrating \emph{SWhisper}'s robustness across challenging real-world scenarios.

\subsection*{E. Extended Ablation Study on Additional Target Models}

Table~\ref{tab:ablation_all} presents an extended ablation study across multiple target models using LLaMA-3.1-8B-Instruct as the surrogate model. This study evaluates the impact of removing individual key components of \emph{SWhisper}. The observed trends are consistent with those reported in Section~6.4, reinforcing the contribution of each component to overall attack performance.

\subsection*{F. Text-Based Attack Performance}

The jailbreak prompts generated by \emph{SWhisper} are text prompts that are converted to audio via text-to-speech (TTS) when attacking speech-driven LLMs. To assess their effectiveness in purely text-based settings, we evaluate the performance of these prompts when directly applied to text-based LLMs, without TTS and subsequent transcription. Table~\ref{tab:without_vioce_transfer} reports the text-only attack performance of \emph{SWhisper} and baseline methods across multiple surrogate and target models.

Results show that existing baseline jailbreak methods generally perform better in text-only attacks, particularly optimization-based approaches such as GCG, which achieve much stronger performance when no voice conversion is involved. However, as these methods are not designed for voice delivery, their performance degrades substantially after TTS and transcription. In contrast, \emph{SWhisper} maintains relatively consistent effectiveness across both text-only and voice-based attacks, with only modest performance differences, highlighting the robustness of its voice-aware design.

\end{document}